\documentclass[usenatbib]{mn2e}

\usepackage{xspace}
\usepackage{amsmath}
\usepackage{epsfig}
\usepackage{defs}
\usepackage{longtable}
\usepackage{multirow}

\title[Seeding High Redshift QSOs]{Seeding High Redshift QSOs by Collisional Runaway in Primordial Star Clusters}

\author[H. Katz et al. ]{Harley Katz$^{1}$\thanks{E-mail: hk380@ast.cam.ac.uk}, Debora Sijacki$^{1}$ and  Martin G. Haehnelt$^{1}$\\
$^1$Institute of Astronomy and Kavli Institute for Cosmology, University of Cambridge, Madingly Road, Cambridge, CB3 0HA\\
}

\begin{document}

\maketitle

\begin{abstract}
We study how runaway stellar collisions in high-redshift, metal-poor star clusters form very
massive stars (VMSs) that can directly collapse to intermediate-mass black holes (IMBHs). We
follow the evolution of a pair of neighbouring high-redshift mini-haloes with high-resolution,
cosmological hydrodynamical zoom-in simulations using the adaptive mesh refinement code
{\small RAMSES} combined with the non-equilibrium chemistry package {\small KROME}. The first collapsing
mini-halo is assumed to enrich the central nuclear star cluster (NSC) of the other to a critical
metallicity, sufficient for Population II (Pop. II) star formation at redshift $z\approx27$. Using the
spatial configuration of the flattened, asymmetrical gas cloud forming in the core of the metal enriched
halo, we set the initial conditions for simulations of an initially non-spherical star
cluster with the direct summation code {\small NBODY6} which are compared to about 2000 {\small NBODY6}
simulations of spherical star clusters for a wide range of star cluster parameters. The final mass
of the VMS that forms depends strongly on the initial mass and initial central density of the
NSC. For the initial central densities suggested by our {\small RAMSES} simulations, VMSs with mass $>400$~M $_{\odot}$  can form in clusters with stellar masses of $\approx10^4$~M$_{\odot}$, and this can increase to well
over 1000~M$_{\odot}$ for more massive and denser clusters. The high probability we find for forming
a VMS in these mini-haloes at such an early cosmic time makes collisional runaway of Pop.
II star clusters a promising channel for producing large numbers of high-redshift IMBHs that
may act as the seeds of supermassive black holes.
\end{abstract}


\section{Introduction}
The most simple path to a black hole is the death of a massive star
of mass M $>8$~M$_{\odot}$; however, these stellar mass black holes cannot
accrete matter fast enough to produce the population of observed
supermassive black holes (SMBHs) at high redshift if growth is
limited to the Eddington rate \citep{Volonteri2003, Volonteri2005, Haiman2006, Lodato2006}.
The problem would be somewhat alleviated if Population III (Pop.
III) stars were hundreds of solar masses as early simulations predicted
\citep{Bromm2001}. Direct collapse of these
stars into black holes of a similar mass could then provide a promising
route to an SMBH if the resulting black hole remnants could
be fed continuously. Recent, higher resolution simulations predict,
however, that the masses of Pop. III stars are significantly lower
due to fragmentation which forms a small stellar association rather
than an individual star. This casts doubt on the possibility that Pop.
III stars produce massive black hole seeds \citep{Greif2011}. Even
if Pop. III stars can reach masses of $\approx1000$~M$_{\odot}$ (see e.g. \citealt{Hirano2014}), there are still major issues regarding the accretion once
the black hole forms which are mainly due to radiative feedback
\citep{Johnson2007,Park2011}.

A different route to massive black hole seeds is the direct collapse
of massive gaseous cores in the centres of atomic cooling haloes
with T$_{\text{vir}}\gtrsim10,000$~K at high redshift. In primordial galaxies that
remain free from molecular hydrogen and metals, the gas cools very
inefficiently at lower temperatures than the atomic cooling limit and
matter will continue to accrete on to the central object at a high rate.
Simulations indicate that the gas may then directly collapse into a
black hole of $\approx10^4-10^6$~M$_{\odot}$ \citep{Loeb1994,Eisenstein1995,Begelman2006,Regan2009,Choi2013}. This mechanism struggles,
however, if the gas can efficiently fragment and cool and it
is essential that the halo remains free from metal, dust, and H$_2$ contamination.

Simulations of the direct collapse scenario therefore invoke a
strong, uniform, Lyman-Werner (LW) background to dissociate H$_2$
\citep{WG2011b}. The most likely environment
to find such a strong LW background is in the vicinity of a much
larger host galaxy which has already undergone significant Pop. II
or Pop. III star formation. There is some debate about the amplitude
of the LW background required to suppress fragmentation (see e.g. \citealt{Shang2010,WG2011,Sugimura2014,Latif2014c,Latif2014b,Regan2014}) especially in the presence of cosmic ray
and X-ray radiation \citep{Inayoshi2011,Inayoshi2014}. Recently, \cite{Regan2014} have shown that if the LW
background is not uniform, the critical value needed to dissociate
the H$_2$ is much higher than that for a uniform background. \cite{Latif2014} suggest that when modelling the radiation spectra of
Pop. II stars correctly and including the impact of X-ray ionization,
the number density of direct collapse black holes decreases
below that required to grow observed high-redshift SMBHs. X-ray
feedback from the initial gas accretion on the black hole may further
limit how massive such an object can grow in a short period
of time \citep{Akyutalp2014}. Thus, it appears prudent to explore
other mechanisms for forming massive black hole seeds at high
redshift.

In this work, we study how stellar collisions in high-redshift,
dense star clusters lead to the runaway growth of a single star. As
stars collide, the mass and radius increase which boosts the probability
for future collisions. This process becomes remarkably unstable,
and analytic work, as well as simulations, demonstrates that,
under the right conditions, runaway stellar collisions can produce
a very massive star (VMS) that may collapse to an intermediate-mass
black hole (IMBH) of $\approx1000$~M$_{\odot}$ \citep{Begelman1978,PZ2004,Freitag2006}. For this
reason, high-redshift nuclear star clusters (NSCs) are very promising
candidates for the formation of IMBHs \citep{Omukai2008}.  \cite{Devecchi2009} used analytical
models to demonstrate how a population of IMBHs might form
in clusters at the centres of high-redshift galaxies. Runaway collisions
in Pop. II star clusters may therefore be key for explaining
the presence of black holes over the entire observed mass
range.

Here, we use a combination of hydrodynamic simulations and direct
summation, $N$-body simulations to model collisional runaway
in dense stellar clusters at high redshift. In Part I (Section 2), we begin
with self-consistent cosmological simulations performed with
the {\small RAMSES} code and identify dense baryonic clumps in protogalaxies
for which we can determine detailed chemo-thermodynamical
properties. In Part II (Section 3), we extract the clumps from the
cosmological simulations and populate them with stars, varying the
stellar initial mass function (IMF) as well as a range of other parameters,
and use this as the initial conditions for direct $N$-body simulations
performed with {\small NBODY6}. Our model is nearly self-consistent,
barring the ability to resolve the formation of individual stars within
the cosmological framework, which is only now becoming possible
\citep{Safranek2014}.

\section{Part I: Cosmological Hydrodynamic Simulations}

\subsection{Set-up of the cosmological simulations}

\subsubsection{Hydrodynamic, gravity, and chemistry solver}
\label{HydroChem}
We use the publicly available adaptive mesh refinement (AMR)
code {\small RAMSES} \citep{Teyssier2002} to follow the detailed hydrodynamics
of the first collapsing objects at high redshift. We have replaced
the default cooling module in {\small RAMSES} with the non-equilibrium
chemistry solver {\small KROME}\footnote{www.kromepackage.org}$^,$\footnote{https://bitbucket.org/tgrassi/krome} \citep{Grassi2014}. {\small KROME} uses the high-order
{\small DLSODES} solver to solve the rate equations for the chemistry
network. We follow the detailed abundances of 12 species: H, e$^-$, H$^+$, H$^-$, He, He$^+$, He$^{++}$, H$_2$, H$_{2}^+$, D, D$^+$, and HD for the reactions
listed in the KROME react\_primordial\_photoH2 network. Cooling
due to metals via line transitions is included at $T<10^4$~K for O I,
C II, Si II, and Fe II. The abundances of these species are pinned
to the hydrogen density in each cell assuming a metallicity which
can change throughout the simulation. This is further described in
Section 2.2.1. We emphasize that we do not assume an amplified LW
background that would dissociate H$_2$ molecules and thus prevent
cooling below $\approx10^4$~K and inhibit gas fragmentation.

\subsubsection{Optimizing resolution and refinement}
\label{Res}
For the purpose of our work, it is important to choose a large
enough cosmological box to have a sufficiently massive halo forming
within. At the same time, we need to choose the maximum
level of refinement of the simulation such that we can resolve high
enough densities for star formation to occur while at the same time
making sure that no numerical fragmentation happens.  \cite{Lada2010} suggest that star formation can begin to occur
at volume number densities $n\gtrsim10^4$~cm$^{-3}$.  \cite{Ceverino2010} argue that the Jeans length, $\lambda_j$, must be resolved
by at least $N_j=7$ cells at the maximum level of refinement, lmax,
to prevent numerical fragmentation. Note that this is higher than
the often used value of $N_j=4$ cells suggested by \cite{Truelove1997}. To study the properties of the birth clouds of NSCs that
form in our simulations, we must evolve the simulations past the
point of first collapse which makes these simulations susceptible to
numerical fragmentation \citep{Robertson2008,Ceverino2010,Prieto2013}. In order
to prevent this, we implement an artificial temperature floor based
on the Jeans criteria at the maximum level of refinement,

\begin{equation}
\label{Tfloor}
\text{T}_{\text{floor}}=\frac{G N_{j}^2 \mu m_{\text p} \rho L^2}{\pi \gamma k_{\text b} 2^{2l_{\text{max}}}}.
\end{equation}

Here, $L$ is the physical length of the box at the redshift of interest, $\gamma$
is the adiabatic index, $N_j$ is the number cells we wish to resolve the
Jeans length with, $\mu$ is the mean molecular weight, $\rho$ is the mass
density of the cell, $k_{\text b}$ is the Boltzmann constant, and $l_{\text{max}}$ is the
maximum level of refinement. We set $N_j=8$, double the number
suggested by \cite{Truelove1997} and slightly larger than the value
suggested by \cite{Ceverino2010}. The minimum temperature of
our simulation is governed by the physical temperature floor set by
the cosmic microwave background (CMB) temperature at a given
redshift. Thus, as long as T$_{\text{floor}}\lesssim2.725(1+z)$~K, we will continue
to accurately resolve the chemical and hydrodynamical properties
of the gas. Inserting T$_{\text{floor}}\lesssim2.725(1+z)$~K, $N_j=8$, $n=10^4$~cm$^{-3}$,
and $\mu=1.22$ into equation (1), we see that we can safely resolve
this density, choosing $L=500$~comoving~kpc~$h^{-1}$ at $z=30$ with $l_{\text{max}}=19$. For these values, T$_{\text{floor}}=4.9$~K  which is far below the
temperature, T$_{\text{CMB}}=84.5$~K, we expect the gas to cool to. At
coarser levels $<l_{\text{max}}$, we refine in order to resolve the Jeans length
by 16 cells, double our choice of $N_j$. In addition to these criteria,
we have also implemented refinement criteria when the number of
dark matter particles per cell becomes greater than 64 as well as
when the baryons in the cell reach the equivalent scaled mass.

We emphasize that our choice of $N_j$ is unlikely to be sufficient
to resolve the turbulent properties of the gas on the scales that we
simulate. \cite{Federrath2011} and \cite{Turk2012} demonstrate
that in order to capture these properties in simulations, the Jeans
length must be resolved by more than $\approx32$ cells especially in the
presence of magnetic fields. As this is not the aim of our work
and we only look to model the general structure and mass of the
birth cloud of a high-redshift primordial star cluster, our choice of
resolution should be sufficient.

\subsubsection{Initial Conditions}
\label{ICs}
We use the software package {\small MUSIC} \citep{Hahn2011} to construct
initial conditions for a collisionless (dark-matter-only) simulation
using second-order Lagrangian perturbations on a uniform grid at
$z=150$ with $256^3$ particles in a 500~kpc~$h^{-1}$ comoving box. For
the cosmological parameters, we assume the most recent values
from \cite{Plank2013} ($h=0.6711$, $\Omega_m=0.3175$, $\Omega_{\Lambda}=0.6825$, $\sigma_8=0.8344$). The transfer function used to generate
the initial conditions was created using {\small CAMB} \citep{Lewis2000}. We use the {\small ROCKSTAR} halo finder \citep{Behroozi2013} to identify the most massive halo in our simulation
at $z=20$ which has a mass of $M_{\text{vir}}=2.48\times10^7$~M$_{\odot}$. We define
a cubic Lagrange region of side 127 comoving kpc to encompass
all of the particles at $z=20$ within a region much larger than the
virial radius of the halo. Multiple dark-matter-only simulations were
run until an atomic cooling halo with $T_{\text{vir}}\gtrsim10^4$~K was identified
at this redshift. An object of this mass is slightly overmassive for
our choice of box size and represents an rms fluctuation of $>4{\sigma}$
indicating that it is a rare halo. Such rare haloes are likely to be
incorporated into the most massive galaxies at lower redshifts and
are thus likely sites to host SMBHs at $z\gtrsim6$ \citep{Sijacki2009,Costa2014}.

Baryons are introduced into the initial conditions using the local
Lagrangian approximation at level 8 on the base grid, and
both the dark matter and the baryons are initially placed at level
11 in the refinement region which gives an effective dark matter
resolution of $2048^3$ particles corresponding to particles of mass
$m_{dm}=1.08$~M$_{\odot} h^{-1}$. The initial level of the refined region was determined
so that the mass of the dark mater particle does not subject
the refined cells to $N$-body heating which can occur when the mass
of the dark matter particle is much greater than the mass of the cells.
Various initial resolutions were tested and it was found that level
11 provides an efficient compromise between particle number and
$N$-body heating effects which were found to be negligible at this
level.

Although the high-resolution dark matter particles are unlikely to
cause spurious heating, it is possible that low-resolution dark matter
particles may infiltrate the refinement region and cause heating
due to their much greater masses. We have checked throughout our
simulation for contamination of low-resolution dark matter particles
and found none of these particles within the virial radius of
the haloes. The maximum level of refinement of our simulation was
defined in Section 2.1.2 to be $l_{\text{max}}=19$ so that the artificial temperature
floor remains less than T$_{\text{CMB}}$ at the redshifts of interest. This
gives us a resolution of $0.95$~comoving~pc~$h^{-1}$, corresponding to a physical resolution of $0.046$~pc at $z=30$.

\subsubsection{Identifying the birth clouds of nuclear star clusters}
We use the clump finder implemented in {\small RAMSES} to identify gravitationally
contracting, bound clumps within the simulations. The
clump finder is sensitive to a density threshold, a mass threshold,
and a relevance threshold\footnote{The relevance threshold is the ratio of the density peak to the maximum saddle density \citep{Bleuler2014}.  Clumps which do not break this threshold are considered noise.}. To identify a clump, we set the minimum
density to be $(1+z)^3$~cm$^{-3}$ which corresponds to $3\times10^4$~cm$^{-3}$ at $z=30$, and the relevance threshold to 1.5. We do not put a constraint
on the minimum mass. We choose the density criteria slightly
higher than the density we wish to resolve in order to minimize the
chance of identifying spurious clumps at the density of interest. This
algorithm is used only to identify the location of clumps within the
simulation. In order to calculate clump properties, we perform a
secondary analysis where we define the outer edge of the clump to
be the radius where the average density drops below $n=10^4$~cm$^{-3}$.
We only consider clumps with $N>N_{j}^3=512$ cells which are all at
the highest level of refinement. For $l_{\text{max}}=19$, the minimum volume of a clump is then 0.05~pc$^3$ at $z=30$ which, assuming a spherical
structure, sets a minimum radius of the clump to be $R_{\text{clump}}>0.23$~pc.
The Arches cluster, the densest known star cluster in the local Universe
with a central density of $\approx10^5$~M$_{\odot}$~pc$^{-3}$, has an inner core
radius of $\approx0.2$~pc which is just about resolved by our simulation
\citep{Espinoza2009}. This suggests that the full
radius of high-density clusters will likely be sufficiently resolved
out to their outer radii by our simulations.

\subsubsection{When to end the simulation?}
\label{when2end}
We end the simulation when high-mass stars are likely to form
as these eventually disrupt the cluster by supernova feedback. We
apply the following criteria to determine when a clump is populated
with high-mass stars. For a given stellar IMF, $dN/dM=\xi(M)$,
the total number of stars above a certain threshold mass, $M_{\text{thresh}}$, is given by $N_*(>M_{\text{thresh}})=\int^{M_{\text{max}}}_{M_{\text{thresh}}}\xi(M)dM$. The main-sequence
lifetime of stars as a function of their mass begins to flatten for
stars with $M\gtrsim40$~$M_{\odot}$. For a metallicity of $Z=10^{-4}Z_{\odot}$, this
corresponds to a main-sequence lifetime of $\approx5$~Myr. Thus, we set
$M_{\text{thresh}}=40$~M$_{\odot}$. We can then compute the mass of the cluster,
$M_{\text{clump}}$ needed to have $N_*(>M_{\text{thresh}})\geq 1$.  For a Salpeter IMF ($\xi(M)\propto M^{-2.35}$), with $M_{\text{min}}=0.1$~M$_{\odot}$ and $M_{\text{max}}=100$~M$_{\odot}$, we find $M_{\text{clump}}=1.3\times10^4$~M$_{\odot}$ for $N_*(> M_{\text{thresh}})=1$. Once the NSC in
the simulation reaches this fiducial mass of $1.3\times10^4$~M$_{\odot}$, we allow
the simulation to run for $t_{\text{lag}}=3.5$~Myr before we extract the clump,
consistent with \cite{Devecchi2009}. Note that the stellar
IMF at high redshift is unknown and the chosen value of $M_{\text{clump}}$ will
change considerably based on the choice of IMF. However, given
that our chosen value of $M_{\text{clump}}$ is much lower than the masses of
observed NSCs, this is a rather conservative assumption.

\subsection{Results from the cosmological simulation}
\label{cosmores}
In order to model the formation of a Pop. II star cluster, we identify a
collapsing cloud of gas in close vicinity to another already collapsed
object. This allows for the first halo to undergo an episode of Pop.
III star formation and enrich the surrounding gas, post supernova, to
a level suitable for forming Pop. II stars. The secondary collapsing
object, however, must be located at a sufficient distance such that the
radiation feedback from the first episode of Pop. III star formation
does not disrupt the collapse
 
\subsubsection{The collapse of two mini-halos in close separation} 
\label{ccsep}
A halo of $M_{\text{vir}}=2.48\times10^7$~M$_{\odot}$ was identified at $z=20$ in the
dark-matter-only simulation with comoving box size 500~kpc~$h^{-1}$.
The member particles were traced back to the initial conditions
where they were centred and the simulation was reinitialized with
baryons. Within this larger halo, two mini-haloes were identified
with a distance such that the first collapsing object can enrich the
second with metals.

At the time of collapse, the first mini-halo has a virial mass
$M_{\text{vir}}=2.7\times10^5$~M$_{\odot}$, a virial radius $R_{\text{vir}}=62$~pc\footnote{The virial radius is calculated assuming that the average density of the object equals $200\rho_{\text{crit}}$.} and a baryon
fraction of 14.2 per cent. This halo begins to collapse at $z=31.6$
and is the first object to collapse in the entire simulation volume.

\begin{figure}
\centerline{\includegraphics[scale=.5,clip,trim=0cm 0cm 2cm 1.3cm]{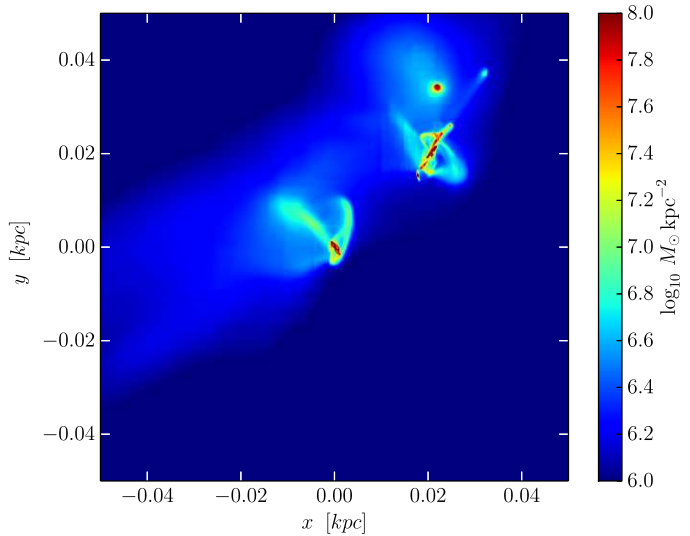}}
\caption{Snapshot showing the surface density in a 100 pc cube of the two mini-halos 8.5 Myr after the collapse of the second object.  The secondary collapsing halo is centered.}
\label{Gascomp}
\end{figure}

If we assume that the average mass of Pop. III stars is 40~M$_{\odot}$ \citep{Hosokawa2011}, the average lifetime of the system prior
to supernova explosion is 3.9 Myr \citep{Schaerer2002}. We model the
chemical enrichment from these first supernovae by implementing a
metallicity floor of $10^{-4}$~$Z_{\odot}$ at $z=30.7$. While this choice of metallicity
floor is arbitrary, a prerequisite for the formation of a Pop. II
star cluster is that the halo be enriched to a metallicity above the
critical metallicity for fragmentation. For dust-free gas, this value
is approximately $10^{-4}$~$Z_{\odot}$ \citep{Schneider2012}. Slightly higher
metallicities are unlikely to significantly affect the hydrodynamics
of the gas at the densities our simulations probe, because as we will
see, the temperature of the gas in the second mini-halo reaches the
CMB temperature floor which prevents further cooling. A slightly
higher metallicity may only accelerate this process.

The simulations of \cite{Ritter2012} model the transport of metals
explicitly from the supernova of Pop. III stars and demonstrate
that the surrounding medium can be enriched to metallicities as
high as $10^{-2}\ Z_{\odot}$. The metal enrichment extends all the way to the
virial radius of the halo in only 8.5 Myr. The outflow is expected to
be bipolar and the dense filaments feeding the halo should be only
minimally enriched. The first mini-halo studied here is less massive
than the halo in \cite{Ritter2012}. We may therefore expect more
mixing along the filaments as the outflow should be less impeded
by gas in the central regions of the halo in our simulation.
 
At $z=28.9$, the second object begins its collapse at a distance
of 117 pc from the centre of the first mini-halo and slightly outside
its virial radius. This occurs 12.9 Myr after the collapse of the
first object. This leaves sufficient time for the first object to form
Pop. III stars and enrich the surrounding material with metals. The
first mini-halo has since grown to $M_{\text{vir}}=8.33\times10^5$~M$_{\odot}$ with $R_{\text{vir}}=97$~pc and a baryon fraction of 16 per cent. The second object
is falling into the potential well of the first mini-halo along a dense
filament (see Fig. 1) and the two objects will eventually merge.

Assuming a population of 40 M$_{\odot}$ Pop. III stars, we can estimate the expected intensity of the LW radiation at the location of the second mini-halo as,
\begin{equation}
J_{21}=10^{21}\frac{{\dot N_{\text{ph}}}h_{\text p}}{4\pi^2r^2},
\end{equation}
where $J_{21}$ is in units of $10^{-21}$erg~cm$^{-2}$~s$^{-1}$~Hz$^{-1}$~sr$^{-1}$, ${\dot N_{\text{ph}}} =\bar{Q}N_{*}f_{\text{esc}} $ is the total number of photons emitted per second which escape the galaxy for a population of $N_{*}$ Pop. III stars with mass of 40~M$_{\odot}$, $\bar{Q}=2.903\times10^{49}$~photons~s$^{-1}$ \citep{Schaerer2002} for a 40~M$_{\odot}$ Pop. III star, $f_{\text{esc}} $ is the escape fraction, and $h_{\text{p}} $ is Planck's constant. Assuming a maximal $f_{\text{esc}} =1.0$ and a separation of 117~pc, we find $J_{21}=9.37N_{*}$.  

\cite{Regan2014} have demonstrated that for an anisotropic
LW source, the critical value of $J_{21}$ needed to completely dissociate
H$_2$ and keep the gas from cooling is $J_{21}\approx10^3$. At a maximal escape
fraction, complete dissociation would require the formation of a
small cluster of roughly 100 Pop. III stars with masses of the order
of 40~M$_{\odot}$ at the centre of the first mini-halo whereas simulations of
these mini-haloes tend to predict small stellar associations of Pop.
III stars \citep{Greif2011}. Our simulations do not include dust
which should be present in the second mini-halo and will help the
gas cool and catalyse the formation of H$_2$. One-zone models predict
a lower critical value of $J_{21}$ needed to disrupt the formation of HD
which is the dominant coolant below $\approx$200\,K \citep{Yoshida2007,WG2011}. This is because
the main formation channel of HD, H$_2$ + D$^+$ $\rightarrow$ HD + H$^+$,
requires the presence of H$_2$ which is lowered by the exposure to
the meta-galactic UV background. Given the low mass of the first
mini-halo and the initial separation of the objects when the secondary
collapses, it is unlikely that the radiative feedback from the
first mini-halo will significantly disrupt the temperature and density
evolution of the second. For these reasons, we do not include a
meta-galactic UV background in our simulation. This approach is
likely to be conservative as even a mild UV background can increase
the accretion rate on to a central NSC \citep{Devecchi2009}.
Denser and more massive NSCs should more efficiently undergo
runaway stellar collisions, and by neglecting a positive contribution
from the UV background we may underestimate the final masses of
the VMSs.

\begin{figure*}
\centerline{\includegraphics[scale=1.1,angle=-90,clip,trim=7.5cm 1.8cm 7.5cm 9.4cm]{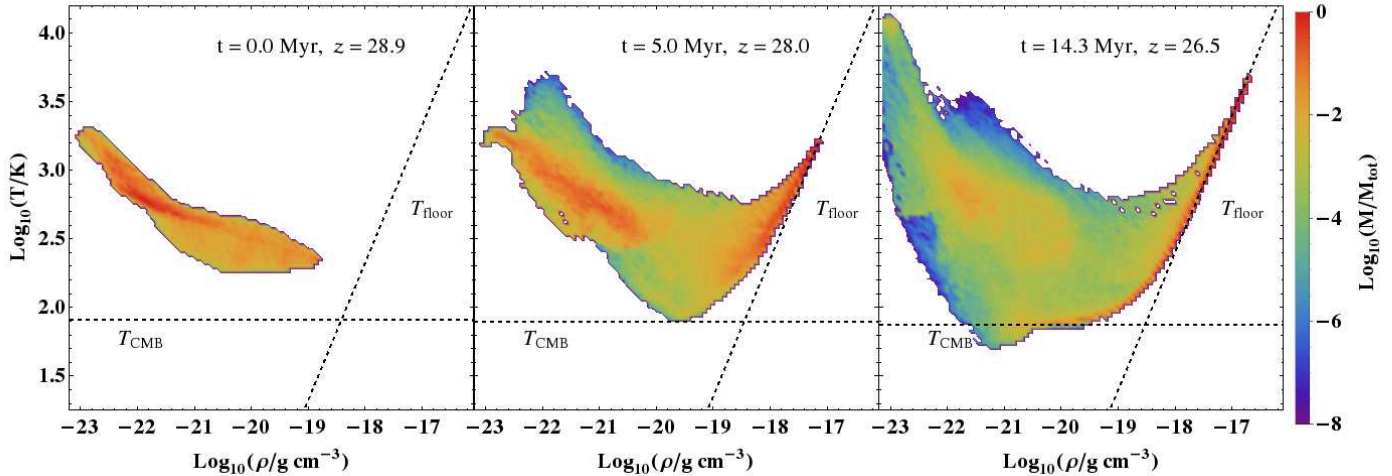}}
\caption{Mass-weighted phase space diagrams of density versus temperature at the initial collapse ({\it left}), 5 Myr after the collapse ({\it middle}), and 14.3 Myr after the collapse ({\it right}) which is the point at which we extract the clump properties from the simulation for input into direct $N$-body calculations.  This includes all gas within 10 pc of the densest cell.  The gas cools to a few hundred kelvin initially due to H$_2$, cooling then reaches the CMB temperature floor due to HD and metal cooling.  The dashed lines represent the CMB temperature floor and the artificial temperature floor as labelled.}
\label{PT:BA}
\end{figure*}

\subsubsection{Evolution of the second mini-halo}
In Fig. 2 (left and middle panels), we plot the mass-weighted phase
space diagram of density versus temperature of the second minihalo
within 10 pc of the densest cell just as it is collapsing at $z=28.9$ and then 5 Myr later at $z=28$. In these early phases of the collapse,
cooling is dominated by H$_2$ lowers the gas temperature and the
mass fraction of HD continues to rise at higher densities. HD and
metals significantly contribute at higher densities until either the
CMB temperature floor or the artificial temperature floor inhibits
further cooling.

In the right-hand panel of Fig. 2, we show the phase space diagram
at 14.3 Myr after the initial collapse. By this point in the simulation,
much of the highest density gas has reached either T$_{\text{CMB}}$ or is
affected by the artificial temperature floor. One-zone models predict
that the ratio $n_{\text{HD}}/n_{\text{H}_2}$ peaks around the maximum density we probe
in this simulation before falling off steeply at higher densities. We
do not probe these very high densities in our simulation and the HD
abundance is forced to remain at a high value even though the gas
should further collapse and lower the HD abundance. Some of the
extra HD can diffuse out of the clump and lower the temperature
of gas. This effect would cause an increase in fragmentation and
decrease the mass of our central clump. With the inclusion of the
CMB temperature floor, this effect is somewhat mitigated because
the gas can only cool to T$_{\text{CMB}}$.

Because the small-scale fragmentation properties of the gas are
subject to numerical resolution effects, we run two additional simulations
in which we vary $l_{\text{max}}$ between 18 and 20 and bracket our
fiducial resolution. In Fig. 3, we plot the mass and density profiles
of the halo computed using log-spaced bins centred on the centre
of mass of the halo for three different resolutions. At the time of
collapse (leftmost column), the mass and density profiles are well
converged among the three different resolutions. By $z=27.7$, a
secondary clump begins to emerge in the higher resolution runs
which appears as a bump in the density profile; however, the mass
profile remains well converged. As the clumps begin to interact, we
see some discrepancy between the different resolution runs in the
very central regions; however, in all cases, the mass profile is well
converged out to $1.5$~pc. In Appendix A, we discuss how the small-scale
differences and the formation of secondary clumps affect our
results.

\begin{figure*}
\centerline{\epsfig{figure=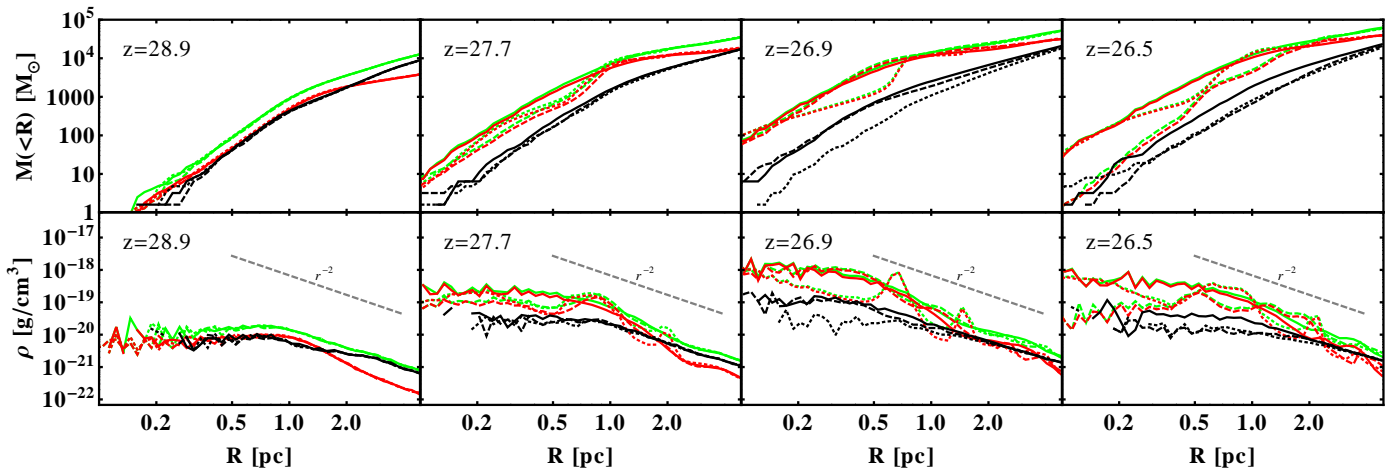,scale=.43}}
\caption{Enclosed mass profiles ({\it top}) and density profiles ({\it bottom}) for three different resolutions as a function of redshift.  The dark matter, gas, and total profiles are represented by black, red, and green lines, respectively.  The solid, dashed, and dotted lines represent $l_{\text{max}}=$ 18, 19, and 20.}
\label{mdlogprof}
\end{figure*}

In Fig. 4, we return to the run with $l_{\text{max}}=19$ and plot the evolution
of the mass and radius of the clump as a function of time since initial
collapse. Both properties increase linearly for $\approx6$~Myr until a second
clump appears in the central regions of the mini-halo. The properties
of the primary clump begin to oscillate as the presence of secondary
fragments increases the average density at larger radii farther away
from the centre. Despite this interaction, we can see in the top panel
of Fig. 4 that the general trend is for the mass to increase. In order
to obtain a smooth function for the mass of the clump as a function
of time, we fit a piecewise linear model to the data points extracted
directly from the simulation and set the break in the function to be
the point where the oscillations begin. We apply this technique for
both the mass and the radius of the clump of interest. Note that the
radius remains relatively constant after the break.

10.8 Myr after the initial collapse, the central clump has grown to
the threshold mass of $1.3\times10^4$~M$_{\odot}$. The average mass accretion on
to the clump is $\dot{M}_{\text{clump}}\approx6.0\times10^{-4}$~M$_{\odot}$~yr$^{-1}$. The average mass
of dark matter within the clump marginally decreases although
this only affects the total mass accretion rate on to the clump by
5 per cent. At 14.3~Myr, which represents 0.8~Myr $+\ t_{\text{lag}}$, the clump
reaches a mass of $1.77\times10^4$~M$_{\odot}$ within a radius of 1.25~pc.

\subsubsection{Internal clump structure}
We assume that the stars form in the highest density regions of
the clump. We identify the densest simulations cells making up a
certain fraction the total mass and define them to be star forming.
For reasons described in Section 3.1.5, we set the star formation
efficiency (SFE), $\epsilon=2/3$. In Fig. 5, we plot the spatial distribution
of these cells as viewed along the axis with the highest magnitude
of angular momentum ($J_z$ axis) for each of the three different resolution
simulations within the star-forming radius of 1.25 pc from
the densest cell at three different times. Tracing this region throughout
the simulation, we see that at the time of initial collapse (top
row of Fig. 5), all three resolutions exhibit a very similar spatial
distribution. 5 Myr later (middle row of Fig. 5), the spatial distributions
of the three simulations still agree reasonably well, but there
is clear evidence that the higher resolution runs collapse further and
we begin to see fragmentation. By 14.3 Myr, the distribution of
the gas at the different refinement levels has completely diverged
although the mass contained within the region remains reasonably
consistent. The two higher resolution simulations have fragmented
into multiple clumps while the $l_{\text{max}}=18$ run contains only one
object. Interactions between the different clumps in the runs with
higher level of refinement have caused a significant deviation in the
dynamics in the central region which makes the structures appear
different. As pointed out by \cite{Regan2014b},
interpreting the results of AMR simulations at different refinement
levels is non-trivial and choosing a refinement level that is either
too high or too low can lead to misinterpretation of the overall
dynamical evolution of collapse simulations. Our simulations indicate
that fragmentation of the central clump is likely and therefore
$l_{\text{max}}=19$ is the minimum resolution required in this study to resolve
the central dynamics and this is the resolution we choose for further
analysis.

We hence use the spatial structure identified in the $l_{\text{max}}=19$ {\small RAMSES} simulation to generate reasonably realistic initial conditions
for {\small NBODY6} simulations of the NSCs. We caution here, however, that
this approach is by no means unambiguous, and we aim to sample
realistic rather than exact initial conditions. The resolution effects on
the transition from the {\small RAMSES} to the {\small NBODY6} simulations are further
discussed in Appendix A where we demonstrate that regardless of
which resolution is chosen between our two high-resolution runs,
the masses of the VMSs which form remain consistent with each
other.

For the run with our fiducial resolution with $l_{\text{max}}=19$, the total
volume of cells which represent the densest $2/3$ of the total gas
mass in the star-forming region is 0.109~pc$^3$. In this simulation, we
identify three distinct regions: a main central massive clump (clump
1), a less massive, off-centred clump (clump 2), and a further less
massive, off-centred clump (clump 3) as indicated in the bottom-middle
panel of Fig. 5. The masses of these clumps are 9094.4~M$_{\odot}$, 835.4~M$_{\odot}$, and 170.4~M$_{\odot}$, respectively. The structure, orientation, and
relative velocities of these clumps are used in Part II of this work
to generate a series of non-spherical initial conditions for direct
N-body simulations.

\begin{figure}
\centerline{\epsfig{figure=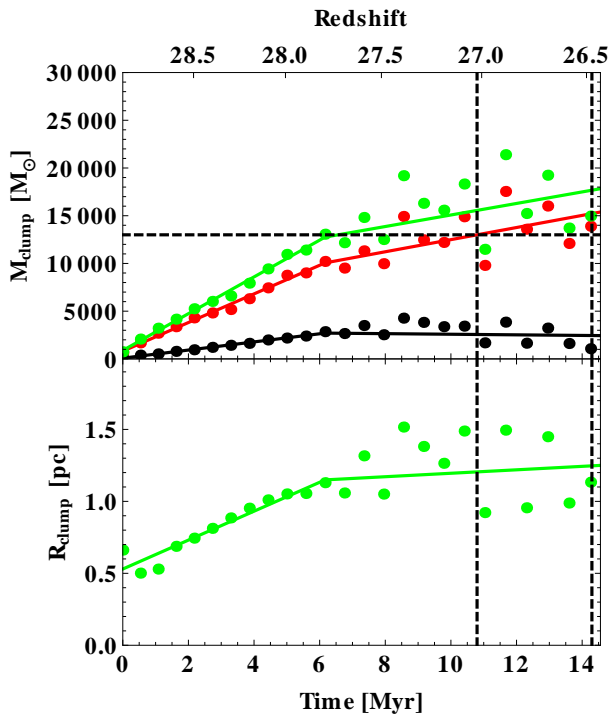,scale=.6}}
\caption{{\it Top.} Mass of the clump in the secondary collapsing object as a function of time since the initial collapse.  The data points represent the  simulations and the solid lines represent  the fitting function described in the text.  The black line shows the dark matter mass, the red line shows the gas mass, and the green line shows the total mass.  The horizontal, dashed, black line represents the fiducial mass at which we expect the cluster to form high mass stars.  The first vertical, dashed, black line is the time at which the clump reaches the fiducial mass and the second vertical, dashed, black line is 3.5 Myr after the first which is the point at which the clump mass is extracted.  {\it Bottom.} Radius of the clump where the average density drops below $10^4$~cm$^{-3}$ as a function of time since the initial collapse.  The dashed, black vertical lines are at the same times as in the top panel and a  piecewise linear fit to the data is also shown.}
\label{cgrow}
\end{figure}

\section{Part II: Direct N-body Simulations}

\subsection{Setup of the {\small NBODY6} simulations}
\subsubsection{From birth cloud to star cluster}
\label{NIIC}
To create initial conditions for the direct N-body simulations, a
minimal bounding ellipsoid enclosing all of the cells for each of
the individual regions shown in the bottom-middle panel of Fig. 5
is computed\footnote{In addition to these three regions, there is a diffuse ring of mass located around the central clump and parallel to the $J_z$ axis.  The total mass in this ring is 5 per cent of the total mass in the main, central clump and we have included it in the mass we list for clump 1 but do not take into account its volume.}. We aim to create a set of star cluster initial conditions
which have the same spatial structure and bulk velocity
properties as the gas bounded by the three ellipsoids. The three
clumps/star clusters will then be evolved together and allowed to
interact. To construct flattened, ellipsoidal star clusters for each individual
clump identified in the birth cloud, we first average the
three primary axis lengths and then generate a spherical star cluster
with this radius with {\small MCLUSTER} \citep{Kupper2011}. For each initial
sphere, the axial ratios are scaled and the positions of individual
stars are rotated and translated according to the properties of the
ellipsoids to reproduce the shapes and orientation of the individual
clumps with respect to each other. The velocities of individual
stars are initialized by choosing a value of the virial parameter $Q$ ($Q=0.5$ represents a virialized cluster and $Q<0.5$ represents dynamically cold clusters) for the initial spherical star cluster prior
to axis scaling and the velocities are rotated to be consistent with
the reorientation of the positions. We do not attempt to scale the
magnitudes of the velocities and only vary $Q$ for the initial spherical
cluster which determines how dynamically cold the initial cluster
is. The bulk velocities of clumps 2 and 3 with respect to clump 1 are
calculated directly from the hydrodynamic simulation and added
to the initial velocities of individual stars of clumps 2 and 3, but
we do not include a net rotation for the central clump which may
delay core collapse. The initial conditions for each of the individual
clumps are then merged into one file to make a single input for
{\small NBODY6}.

The initial density profile of the spherical star clusters, prior to
axis scaling, is varied between a Plummer model \citep{Plummer1911}
and a constant-density model. We further use the fractal dimension
option of {\small MCLUSTER} to model inhomogeneous systems where
stars are sub-clustered to account for the spatial inhomogeneity in
real star-forming systems. $D=3.0$ represents a cluster with no
fractalization and $D=1.6$ represents a very clumpy distribution
\citep{Goodwin2004}. We should note here that for both
the Plummer and the fractal models not all stars in the initial conditions
are placed within the bounding ellipsoid. Higher values of
$D$ effectively increase the initial volume that contains the stars and
therefore lower the average initial densities.

The dark matter and gas not converted to stars are modelled as
an external potential with a Plummer sphere and the virial radius is
set to be 1.25~pc, matching that of the cosmological, hydrodynamic
simulation. The initial mass of the external potential is 7524.7~M$_{\odot}$ and this mass increases at a rate of $6.04\times10^{-4}$~M$_{\odot}$~yr$^{-1}$  over the
course of the simulations as measured from the {\small RAMSES} simulation to
represent the gas and dark matter accretion on to the NSC. Accretion
at this rather low rate affects the dynamics of the star cluster very
little, but is, nevertheless, included for completeness. Note that
accretion at higher rates could play an important role in the evolution
of the cluster.

\subsubsection{Gravity solver}
The evolution of these embedded star clusters and the formation of
VMSs from runaway stellar collisions are modelled with the GPUaccelerated
version of {\small NBODY6} \citep{Aarseth1999,Nitadori2012}. The minimum energy conservation requirement is set so $\Delta E/E\leq10^{-3}$. Because the remaining gas and dark matter are
modelled as a background potential which is variable as a function
of the gas accretion and expulsion rates, energy is not strictly
conserved.

\subsubsection{Stellar evolution and metallicity effects}
\label{metal}
Stellar evolution for individual stars is modelled with the {\small SSE} and
{\small BSE} packages \citep{Hurley2000,Hurley2002}
built into {\small NBODY6}. All stars begin on the zero-age main sequence
(ZAMS).

The lifetimes of the most massive stars are reasonably constant at
M $>40$~M$_{\odot}$, and range from $\approx3.5-5$~Myr with decreasing mass.
These values are rather independent of metallicity. At this epoch in
the cluster lifetime, the most massive stars begin to undergo supernovae,
a process which is not modelled in our simulations. For this
reason, all simulations are stopped at 3.5~Myr when the first massive
star evolves off the main sequence. Contrary to main-sequence lifetimes,
radii of stars are heavily dependent on metallicity and only
converge for lower mass stars. The stellar evolution package native
to {\small NBODY6} does not inherently sample the metallicity of our cluster.

The models of \cite{Baraffe2001} predict that
stars with mass $M\approx100$~M$_{\odot}$ and metallicity $Z=10^{-4}Z_{\odot}$ have
radii of roughly half that of the lowest metallicity native to {\small NBODY6}.
We apply a correction term to the radii to account for this. At
$M <10$~M$_{\odot}$, all radii are as computed for the lowest metallicity
available in {\small NBODY6} and for $M >10$~M$_{\odot}$, a linear interpolation is
applied so that 100~M$_{\odot}$ stars have a radius half that predicted by
the stellar evolution packages.

\begin{figure*}
\centerline{\includegraphics[scale=.42,clip,trim=1.3cm 0.5cm 1cm 1cm]{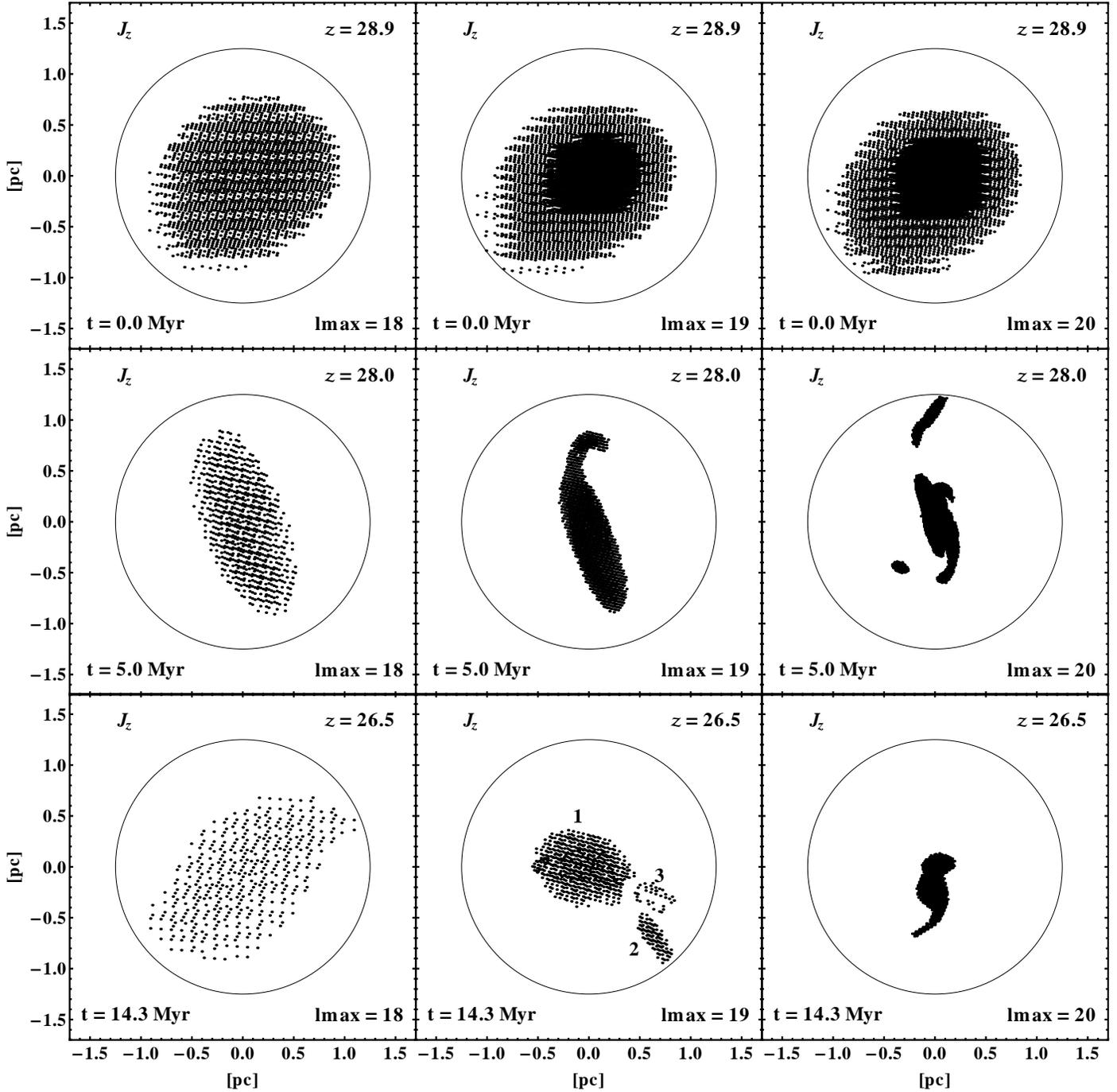}}
\caption{View along the $J_{z}$ axis of the densest cells which represent 2/3 of the total mass within a 1.25 pc radius of the densest cell for the three different resolutions (columns) and at three different times (rows).  The top row shows the initial collapse, the middle row is 5 Myr after collapse, and the bottom row is 14.3 Myr after the collapse, the point at which we extract the clump properties.  Points in the bottom row represent the corresponding star forming region of the clump.  The three clumps  labelled 1, 2, and 3  in the bottom middle panel represent the three clumps which were extracted from the $l_{\text{max}}=19$ simulation and used to create non spherical initial conditions for simulation with {\small NBODY6}.  The circle indicates a radius of 1.25 pc.}
\label{ellpfo}
\end{figure*}

\cite{Glebbeek2009} demonstrated that at high metallicity, stellar
winds become the dominant mode of mass-loss from a star and
can significantly limit the final remnant mass of a collision product.
We have implemented the mass-loss rates of \cite{Vink1999,Vink2000,Vink2001} and apply them to the collisionally produced
stars with M~$>100$~M$_{\odot}$. The strength of the wind scales with
the metallicity. Lower metallicity stars undergo far less mass-loss
than their higher metallicity counterparts. For the metallicity considered
here, stellar winds are inefficient at decreasing the mass of
the VMS over the main-sequence lifetimes of these stars. However,
real NSCs are likely to exhibit a range in metallicities and this will
become important if the metallicity of the cluster is significantly
increased.

Another consequence of the decreased wind strength is the inability
of the stars to unbind the gas from the cluster. This may decrease
the number density of stars throughout the cluster. Using models
from {\small STARBURST99}, \citep{Leitherer1999}, for a Salpeter stellar IMF
with a maximum mass of 100~M$_{\odot}$ and an instantaneous starburst,
we can calculate the integrated mechanical luminosity of our star
cluster by scaling the mass and $Z^{0.5}$. Our simple model assumes that
the mechanical luminosity is dominated by winds rather than radiation.
By comparing this energy input to the binding energy of the
cluster, we find that the energy input from the stars only becomes
comparable to the binding energy of the cluster at $\approx3.5$~Myr which
is the lifetime of the most massive stars in the cluster and the point
at which we stop the simulations. This calculation also assumes that
all of the mechanical luminosity couples to the gas efficiently. We
can, therefore, safely neglect gas expulsion in our simulations.

\subsubsection{Treatment of stellar collisions}
\label{toc}
As the simulations begin with all stars on the ZAMS and are truncated
after 3.5~Myr, the only collisions which can occur are those
between two main-sequence stars. A sticky sphere approximation
is used so that if the distance between the centres of the two stars
is less than the sum of the radii, it is assumed that the stars have
merged. All of the stars in the simulations are on the main sequence
and we assume that when stars collide, the remnant is also a main-sequence
star. The new star is assumed to be well mixed and the new
lifetime is given to be consistent with \cite{Tout1997}. This results
in a slight rejuvenation of the lifetime. If the collision product has
M~$>100$~M$_{\odot}$, the resulting evolution is treated as a 100~M$_{\odot}$ star.

Smoothed particle hydrodynamic simulations which have studied
the merges of main-sequence stars have shown that not all of the
mass that enters a collision is necessarily retained. The mass ratio
of the stars, as well as the orientation of the collision, ranging from
head on to grazing, influences the amount of mass that is lost \citep{Trac2007,Dale2006}. \cite{Glebbeek2013}
demonstrated that the mass-loss is also slightly dependent on the
types of stars which merge. Since all stars in our simulations are on
the main sequence, we adopt approximated mass-loss rates consistent
with the high-mass half-age main-sequence merger models of
\cite{Glebbeek2013} as follows:
\begin{equation}
dM=\min\left[0.062\frac{M_2}{0.7M_1},0.062\right](M_1+M_2),
\end{equation}
where $M_1$ is the mass of the primary and $M_2$ is the mass of the
secondary. The mass-loss is roughly constant for all mass ratios
with $M_2/M_1>0.7$. This is enforced in our equation. We only
calculate this mass-loss when the stars collide almost head-on such
that the distance between the two centres is less than half the sum of
the two radii of the stars and when the orbital kinetic energy of the
secondary star just prior to the collision is greater than the binding
energy

\begin{figure*}
\centerline{\epsfig{figure=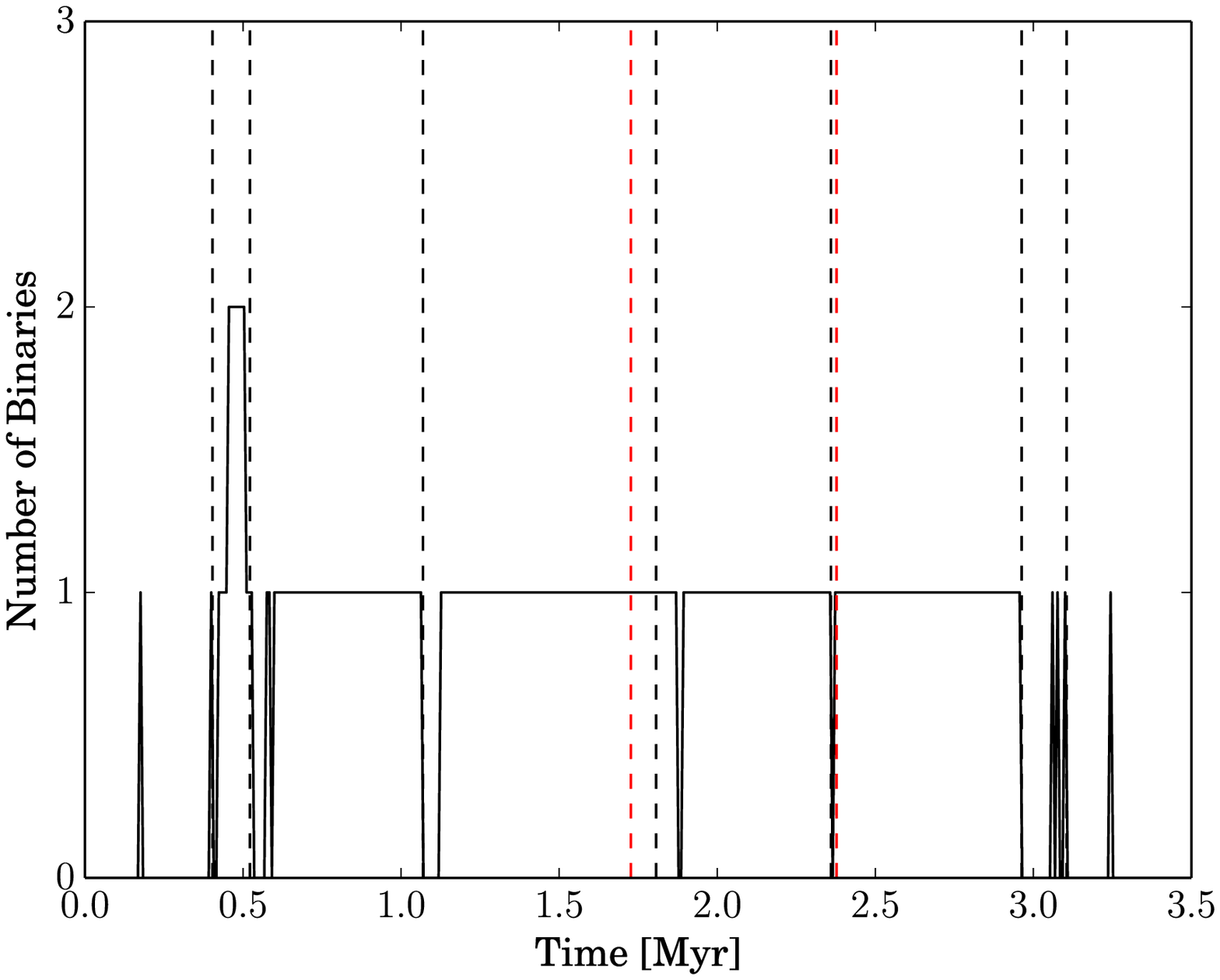,scale=.45}\epsfig{figure=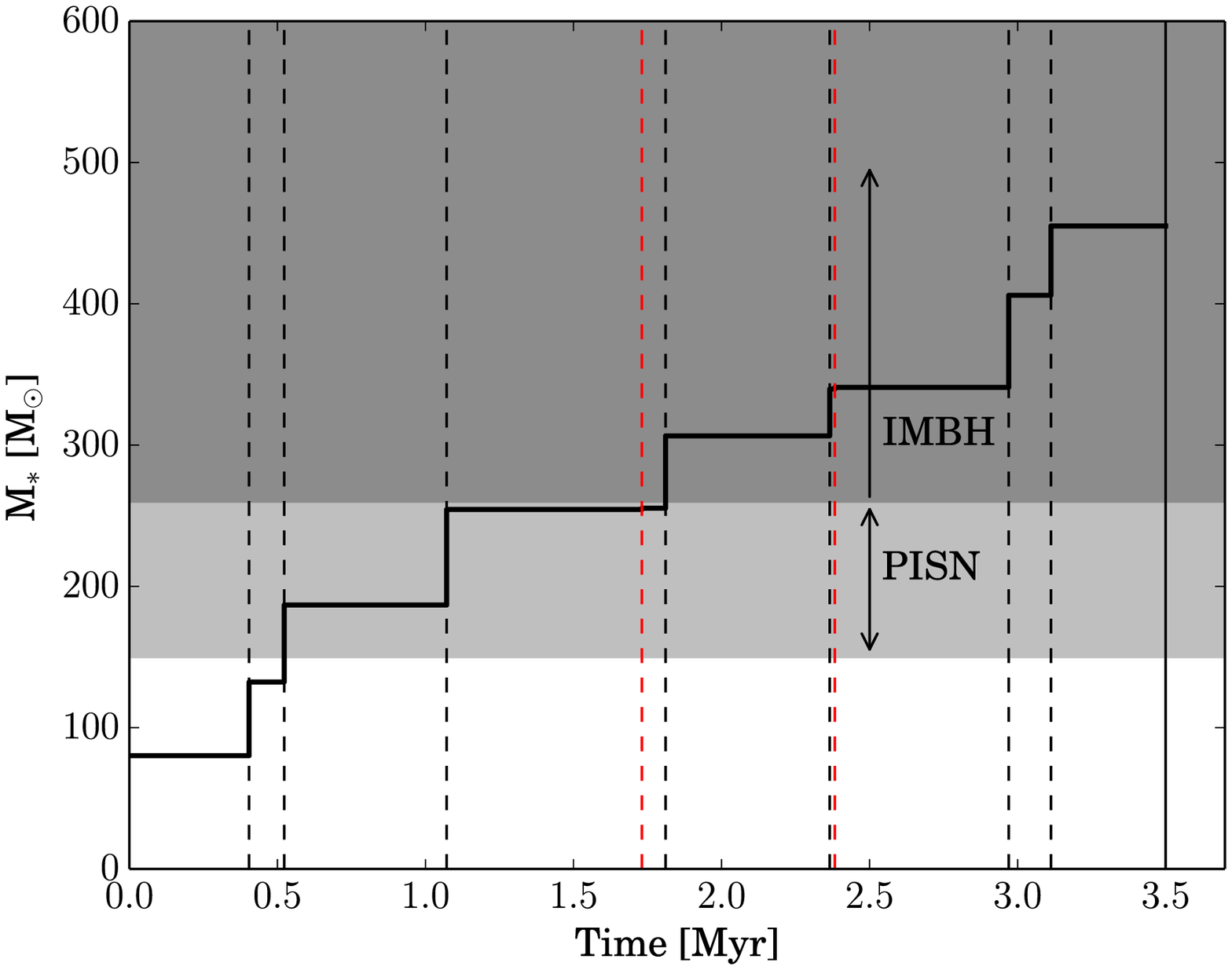,scale=.45}}
\caption{{\it Left.} Number of binaries as a function of time for a representative {\small NBODY6} simulation with the top-heave Saltpeter IMF, with no initial binaries or primordial mass segregation, that results in a VMS with a mass of $M_{\text{VMS}}=455.1$~M$_{\odot}$ after 9 separate collisions.  The time at which the VMS undergoes a collision/merger is indicated with a vertical line where black represents a binary collision and red represents a hyperbolic collision.  {\it Right.} The mass of the collisional runaway as a function of time for the same simulation.  The solid vertical line is the lifetime of the most massive stars in the cluster and the point at which the simulation is stopped.  The light and dark gray regions represent the mass ranges where PISNs and IMBHs potentially form.  In each panel, the time at which the VMS underwent a collision/merger is indicated with a dashed vertical line where black represents a binary collision and red represents a hyperbolic collision.}
\label{bincoll}
\end{figure*}

\subsubsection{Star formation efficiency}
\label{FCC}
In the local Universe the SFE, $\epsilon=M_*/(M_*+M_g)$, is $\approx10-30$ per cent \citep{Lada2003}. However, we do not know whether this is applicable to the high-redshift Universe where the environment
is very different. \cite{Dib2011} have demonstrated that the SFE
increases exponentially with decreasing metallicity with no relation
to the mass of the birth cloud. The metallicity studied in their work
is three orders of magnitude greater than the metallicity floor of
$10^{-4}Z_{\odot}$ used in this study which may suggest that the SFE in the
NSC forming in our simulated galaxy could have $\epsilon\gg35$ per cent.
Simulations of \cite{Pfalzner2013} show that, in order to
reproduce characteristics of local compact clusters of similar mass
to the NSC that forms in our simulation, SFEs of $60-70$ per cent need
to be assumed. This finding is further supported by the simulations
of \cite{Fujii2014} which demonstrate that a local SFE of more than
50 per cent is needed for the formation of young massive clusters
which have properties similar to such objects in the local Universe.
Both observations and the simulation results discussed suggest that
it is indeed appropriate to assume a rather high SFE in our star
cluster simulations.

For our fiducial model, we adopt $\epsilon=2/3$ which is defined at the
point at which we extract the clump properties in the simulation.
We should note, however, that the SFE is dependent on how the
edge of the clump is defined. This is not necessarily consistent
between simulations and observations. We have chosen a fiducial
density threshold of $10^4$~cm$^{-3}$ which is higher than the densities at
the edges of many local molecular clouds. Since observations may
probe lower densities, the SFE which we quote will appear higher
than the true efficiency if a fair comparison would be made between
observations and the simulation. For example, if we consider a
sphere of radius 10~pc around the densest cell and consider only
gas with $\rho\geq10^2$~H~cm$^{-3}$, the effective SFE would be equivalent to
22 per cent.

Because mass is accreting on to the NSC, the SFE calculated
at the beginning of the simulation is higher than it would be if
calculated just prior to the most massive stars going supernova.
With the mass accretion rate of $\dot{M}_{\text{clump}}=6.0\times10^{-4}$~M$_{\odot}$~yr$^{-1}$ taken from our {\small RAMSES} simulation, the effective SFE at the end of
the $N$-body simulation drops to 57 per cent, 10 per cent lower than
the initial value. The impact of varying the SFE on our results is
further discussed in Appendix B where we relax the assumption of $\epsilon=2/3$.

\subsubsection{Possible outcomes of runaway collisions}
There are three possible outcomes of the cluster evolution that we are
interested in identifying: (1) when collisional runaway results in the
formation of a VMS with M~$>260$~M$_{\odot}$, (2) when stellar collisions
result in the formation of a pair-instability supernova (PISN) which
occurs when mergers produce a star with $150<$~M~$<260$~M$_{\odot}$,
and (3) when collisions do not lead to efficient runaway growth and
no star exceeds the PISN mass threshold. In order to sample the
probability of each outcome, we generate multiple realizations of
each set of initial conditions.

The runaway collision process begins when the high-mass stars
sink to the centre of the cluster and dynamically form binaries.
Encounters with other stars perturb these binaries by either three-
(or many-) body scattering or by binary exchanges. The semimajor
axis of the dominant binary continues to shrink as the system loses
energy due to these encounters and if the eccentricity becomes high
enough, the two stars merge. This process of binary capture followed
by a merge repeats until the core evolves to sufficiently low
density or the supernovae from the first massive stars disrupt the
cluster. To demonstrate this process in practice, we show in Fig. 6
the number of binaries as a function of time as well as the mass evolution
of the collisional runaway star for a representative star cluster
simulation which forms a VMS with a mass of $M_{\text{VMS}}=455.1$~M$_{\odot}$
after nine separate collisions. We indicate the times of a collision
and see that most of these coincide with the points at which the
number of binaries changes. The two specific instances where the
collision time is not related to a change in number of binaries occur
when the VMS undergoes a hyperbolic collision. The masses of
the secondary objects in these types of collision tend to be small
with the average secondary mass of the hyperbolic collision being
1.05~M$_{\odot}$ compared to the average mass of the secondary star in
binary collisions being 56.2~M$_{\odot}$. Hence, the hyperbolic collisions
tend to contribute negligible amounts of mass to the overall mass
of the VMS.

\subsection{Results of the direct N-body simulations}
\label{NB6res}

\subsubsection{Non-spherical N-body simulations}
\label{nonideal}

\begin{figure*}
\centerline{\includegraphics[scale=0.36,clip,trim=0cm 0cm 1.5cm 0cm]{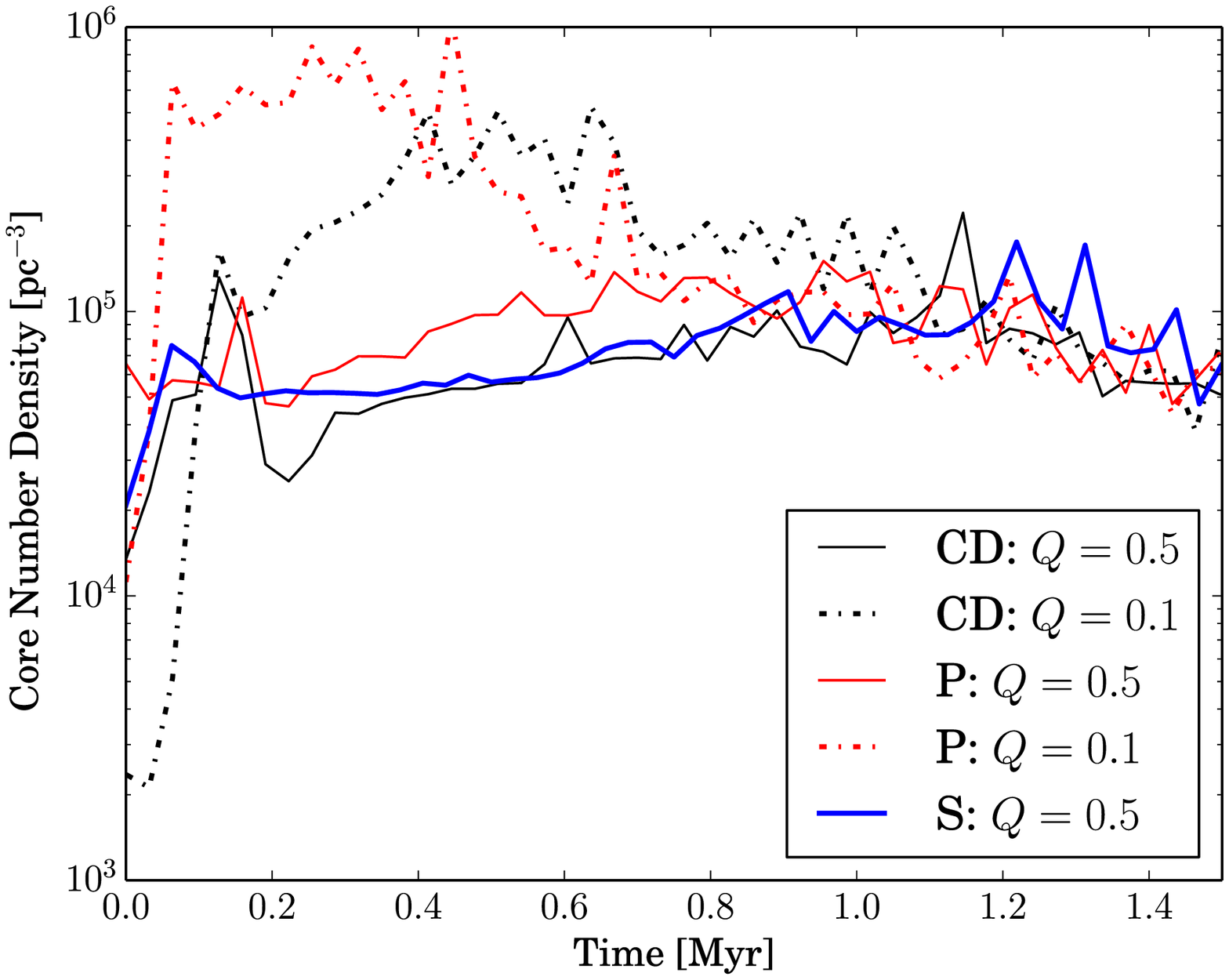}\includegraphics[scale=0.37,clip,trim=1.5cm 0cm 1.5cm 0cm]{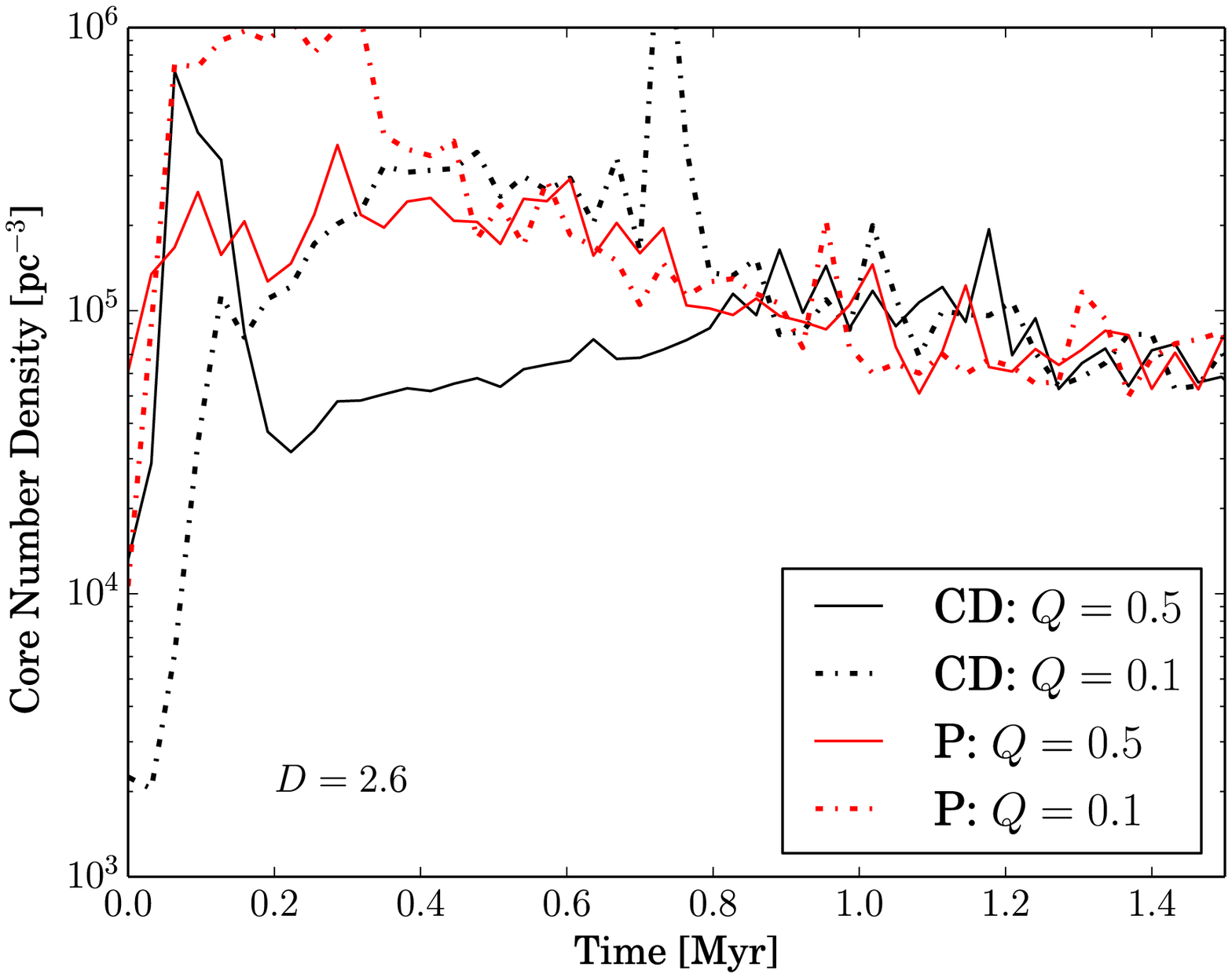}\includegraphics[scale=0.37,clip,trim=1.5cm 0cm 1.5cm 0cm]{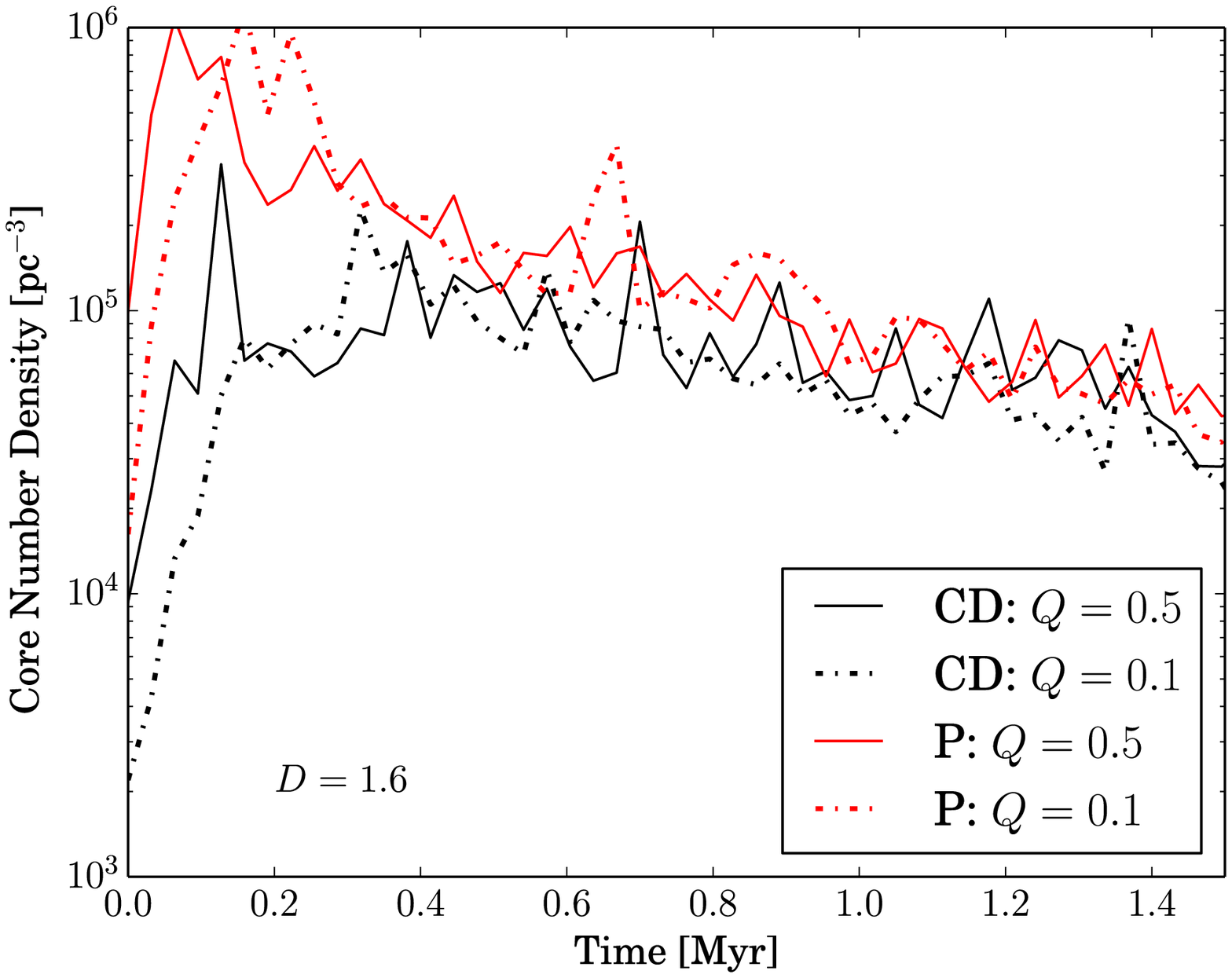}}
\caption{Initial evolution of the number density of the core as a function of time for models with $D=3.0$ (left), $D=2.6$ (middle), and $D=1.6$ (right).  Red lines represent models that were initialized with Plummer potentials (P) while black lines started at constant density (CD).  Initial $Q=0.5$ and $Q=0.1$ are indicated by the solid and dot-dashed lines, respectively.  The blue line in the left panel shows the evolution of a comparison spherical model (S) with $D=3.0$, $Q=0.5$, and $f_c=0.5$.}
\label{ndcoreNI}
\end{figure*}

We begin by studying the non-spherical star clusters which are set up
to reproduce the mass distribution obtained from the cosmological
simulations

\paragraph{Varying the density profile, $Q$ and $D$} 
In Table 1, we list
the initial conditions of all of the models sampled in this section as
well as the results. We fix the stellar IMF to a top-heavy\footnote{By ``top-heavy" we mean that more mass is locked in high mass stars for this IMF compared to another we sample which has the same slope but a lower $M_{min}$.  This should not be confused with an IMF where most of the mass is in high mass stars.} Salpeter
IMF (TH\_Salp) such that $M_{\text{min}}=1$~M$_{\odot}$, $M_{\text{max}}=100$~M$_{\odot}$ and $\alpha=-2.35$. For each of these models, 50 realizations have been
run. The fraction of models which produce a VMS does not exceed
12 per cent and the most massive VMSs have masses in the range
275-410~M$_{\odot}$. The average number of collisions for these models is
rather low. This prohibits collisional runaway in the majority of realizations.
In Fig. 7, we plot the core number density\footnote{The core radius is computed following \cite{Casertano1985} where the density at each particle is calculated using the five nearest neighbours.} for this initial
set of models and see that it increases extremely quickly within the
first 0.25~Myr and slowly decreases thereafter. Even models which
do not form VMSs exhibit this generic behaviour. This number
density rarely peaks above $10^6$~pc$^{-3}$ and the final number density
always converges to values of $\approx2\times10^4$~pc$^{-3}$ at $t=3.5$~Myr. Differences
in the initial degree of fractalization have a very small
effect on the core number density. The total number of collisions
increases in the models with lower values of $D$. However, this does
not affect the fraction of models which produce VMSs. This increase
is likely to be due to a higher initial local number density
in the more fractal models. There is a clear difference in the peak
core number density between models which are initialized with a
Plummer profile compared to those initialized with constant density.
The Plummer models start with higher core number densities and
also reach higher densities. Despite this, the models all converge
within $\approx1$~Myr to a core number density of $\approx10^5$~pc$^{-3}$. The models
which were initialized with lower $Q$ also reach higher densities than
their counterparts. This is expected because the colder clusters are
more susceptible to the initial collapse. The increase in number of
collisions for the coldest clusters is very marginal (see Table 1) and
once again, there is no increase in the fraction of VMSs that are
produced.

\begin{table*}
\centering
\begin{tabular}{@{}lcccccccccccc@{}}
\multicolumn{12}{c}{{\bf Non-Spherical Models}} \\
\hline
\multicolumn{5}{l}{{\bf Initial Conditions}} & \multicolumn{8}{l}{{\bf Results}} \\
\hline
$\rho_{\text{init}}$ & IMF & $Q$ & $D$ & $S$ &$\bar{N}_{\text{coll}}$&$f_{\text{VMS}}$&$f_{\text{PISN}}$&$f_{\text{NE}}$&$\bar{M}_{\text{seed}}$&$M_{\text{VMS,max}}$&$\bar{M}_{\text{VMS}}$&$M_{\text{SMBH}}$\\
& & & & & & & & & M$_{\odot}$ & M$_{\odot}$ & M$_{\odot}$ & 10$^9$~M$_{\odot}$ \\
\hline
P & TH\_Salp & 0.5 & 3.0 & 0.0 & 2.20 & $8.0\pm4.0$ & $32.0\pm8.0$ & $60.0\pm11.0$ & 83.9 & 323.1 & 297.1 & 2.98\\
P & TH\_Salp & 0.3 & 3.0 & 0.0 & 1.84 & $8.0\pm4.0$ & $40.0\pm8.9$ & $52.0\pm10.2$ & 80.9 & 369.9 & 305.3 & 3.06\\
P & TH\_Salp & 0.1 & 3.0 & 0.0 & 1.88 & $4.0\pm2.8$ & $46.0\pm9.6$ & $50.0\pm10.0$ & 82.1 & 303.0 & 297.4 & 2.98\\
P & TH\_Salp & 0.5 & 2.6 & 0.0 & 1.82 & $6.0\pm3.5$ & $24.0\pm6.9$ & $70.0\pm11.8$ & 91.0 & 308.6 & 306.5 & 3.08\\
P & TH\_Salp & 0.3 & 2.6 & 0.0 & 2.04 & $4.0\pm2.8$ & $40.0\pm8.9$ & $56.0\pm10.6$ & 89.9 & 336.5 & 309.1 & 3.10\\
P & TH\_Salp & 0.1 & 2.6 & 0.0 & 2.20 & $8.0\pm4.0$ & $38.0\pm8.7$ & $54.0\pm10.4$ & 71.6 & 370.5 & 332.5 & 3.34\\
P & TH\_Salp & 0.5 & 1.6 & 0.0 & 3.96 & $8.0\pm4.0$ & $38.0\pm8.7$ & $54.0\pm10.4$ & 82.2 & 398.4 & 302.6 & 3.04\\
P & TH\_Salp & 0.3 & 1.6 & 0.0 & 3.34 & $2.0\pm2.0$ & $52.0\pm10.2$ & $46.0\pm9.6$ & 89.2 & 312.6 & 312.6 & 3.14\\
P & TH\_Salp & 0.1 & 1.6 & 0.0 & 4.36 & $10.0\pm4.5$ & $30.0\pm7.7$ & $60.0\pm11.0$ & 73.5 & 356.7 & 316.7 & 3.18\\
P & TH\_Salp & 0.5 & 3.0 & 1.0 & 1.78 & $6.0\pm3.5$ & $32.0\pm8.0$ & $62.0\pm11.1$ & 80.1 & 338.8 & 303.6 & 3.05\\
P & TH\_Salp & 0.3 & 3.0 & 1.0 & 1.80 & $4.0\pm2.8$ & $48.0\pm9.8$ & $48.0\pm9.8$ & 82.6 & 300.6 & 287.5 & 2.88\\
P & TH\_Salp & 0.1 & 3.0 & 1.0 & 1.74 & $12.0\pm4.9$ & $30.0\pm7.7$ & $58.0\pm10.8$ & 88.2 & 339.7 & 326.0 & 3.27\\
CD & TH\_Salp & 0.5 & 3.0 & 0.0 & 1.56 & $6.0\pm3.5$ & $32.0\pm8.0$ & $62.0\pm11.1$ & 89.4 & 302.8 & 288.4 & 2.89\\
CD & TH\_Salp & 0.3 & 3.0 & 0.0 & 1.98 & $4.0\pm2.8$ & $36.0\pm8.5$ & $60.0\pm11.0$ & 74.0 & 275.5 & 268.2 & 2.69\\
CD & TH\_Salp & 0.1 & 3.0 & 0.0 & 2.06 & $10.0\pm4.5$ & $40.0\pm8.9$ & $50.0\pm10.0$ & 77.8 & 411.5 & 305.7 & 3.07\\
CD & TH\_Salp & 0.5 & 2.6 & 0.0 & 2.22 & $4.0\pm2.8$ & $38.0\pm8.7$ & $58.0\pm10.8$ & 76.1 & 320.5 & 310.0 & 3.11\\
CD & TH\_Salp & 0.3 & 2.6 & 0.0 & 1.56 & $0.0\pm0.0$ & $38.0\pm8.7$ & $62.0\pm11.1$ & - & - & - & - \\
CD & TH\_Salp & 0.1 & 2.6 & 0.0 & 2.04 & $10.0\pm4.5$ & $46.0\pm9.6$ & $44.0\pm9.4$ & 86.4 & 332.5 & 302.5 & 3.04\\
CD & TH\_Salp & 0.5 & 1.6 & 0.0 & 7.90 & $4.0\pm2.8$ & $42.0\pm9.2$ & $54.0\pm10.4$ & 81.3 & 310.5 & 289.1 & 2.90\\
CD & TH\_Salp & 0.3 & 1.6 & 0.0 & 7.12 & $8.0\pm4.0$ & $36.0\pm8.5$ & $56.0\pm10.6$ & 82.7 & 428.6 & 321.1 & 3.22\\
CD & TH\_Salp & 0.1 & 1.6 & 0.0 & 8.22 & $2.0\pm2.0$ & $48.0\pm9.8$ & $50.0\pm10.0$ & 54.7 & 278.8 & 278.8 & 2.80\\
CD & TH\_Salp & 0.5 & 3.0 & 1.0 & 1.80 & $2.0\pm2.0$ & $34.0\pm8.2$ & $64.0\pm11.3$ & 99.2 & 358.0 & 358.0 & 3.59\\
CD & TH\_Salp & 0.3 & 3.0 & 1.0 & 2.20 & $10.0\pm4.5$ & $44.0\pm9.4$ & $46.0\pm9.6$ & 80.7 & 355.5 & 302.0 & 3.03\\
CD & TH\_Salp & 0.1 & 3.0 & 1.0 & 2.00 & $8.0\pm4.0$ & $48.0\pm9.8$ & $44.0\pm9.4$  & 94.9 & 320.2 & 295.4 & 2.96\\
\hline
\end{tabular}
\caption{$\rho_{\text{init}}$: the initial density profile of the cluster.  ``P" represents a Plummer sphere and ``CD" represents a constant density model.  IMF: the initial stellar IMF of the models.  $Q$: virial ratio of the initial spherical star clusters.  $Q=0.5$ represents a spherical star cluster in virial equilibrium and $Q<0.5$ represents a dynamically colder system.  $D$: fractal dimension of the cluster.  $D=3.0$ is a smooth cluster and $D=1.6$ represents a very fractal distribution.  $S$: initial degree of mass segregation in the cluster.  $S=0$ represents a cluster without mass segregation and $S=1$ is a fully segregated cluster.  $\bar{N}_{\text{coll}}$: average number of collisions over all of the realizations.  $f_{\text{VMS}}$, $f_{\text{PISN}}$, $f_{\text{NE}}$: percentage of realizations which produce a VMS, PISN, or neither respectively.  Error bars on these values assume Poisson statistics.  $\bar{M}_{\text{seed}}$: average mass of the star which seeded the collisional runaway for simulations which produced a VMS.  $M_{\text{VMS,max}}$: mass of the most massive VMS which formed in all of the realizations.  $\bar{M}_{\text{VMS}}$: average mass of the VMSs which form in the simulations which produced VMSs.  $M_{\text{SMBH}}$: the average mass of a SMBH at $z=6$ assuming Eddington limited accretion with a radiative efficiency of $\epsilon=0.1$.  For all models, 50 realizations are simulated.}
\label{tab:nonideal}
\end{table*}

\paragraph{Primordial mass segregation} 
Many observations suggest
that star clusters may form with a high degree of mass segregation
(see e.g.  \citealt{Gouliermis2004} and \citealt{Chen2007}).
The simulations of \cite{Moeckel2011} demonstrate that during
the accretion phase of gas on to protostars, mass segregation
naturally occurs owning to the distribution of masses of the protostars.
We do not simulate this phase but, if this occurs before
the stars evolve to the ZAMS, our simulations should be initialized
with some degree of mass segregation. Using the algorithm
of \cite{Baumgardt2008} which produces clusters
in virial equilibrium for chosen degrees of mass segregation
(denoted $S=0$ for a primordially unsegregated cluster and $S=1$
for a fully mass segregated cluster), we introduce mass segregation
into the initial spherical clusters prior to scaling the axial ratios and
adjusting the positions and velocities. However, the introduction
of mass segregation into the models has a very minimal effect on
the results of these simulations (see Table 1). The mean number of
collisions remains roughly consistent with models without primordial
mass segregation and there is a tendency for primordially mass
segregated models to produce slightly more PISNs. The number
of VMSs that form remains the same as do the average masses of
the VMSs. Regardless of many assumptions made to produce the
initial conditions, the masses of the VMSs are robust but $f_{\text{VMS}}$ will
change with initial central density and this is further discussed in
Appendix A where we use the non-spherical initial conditions taken
from the $l_{\text{max}}=20$ cosmological simulation to study the formation
of collisional runaway in a denser cluster.

\subsubsection{Spherical N-body simulations}
\label{ideal}
Our simulations taking into account the flattened and asymmetric
spatial distribution of the gas in the central regions of the galaxy in
the cosmological hydrodynamic simulation have succeeded in producing
VMSs which may directly collapse into IMBHs. The simulated
star clusters become spherical within a few hundred thousand
years. However, note that this time-scale is dependent on the initial
rotation within the sub-clumps. This has not been included in our
simulations. Rotation tends to delay core collapse and should prolong
the phase where the star cluster is in an asymmetric flattened
state. While the clusters do become spherical rather quickly, it is
important to note that the initial dynamics, in particular, differ between
the non-spherical and a corresponding spherical cluster. The
evolution of the core number density of a spherical model which
has a similar initial central density follows closely the non-spherical
models with $D=3.0$ and $Q=0.5$ except for an initial spike which
is due to the presence of a secondary clump in the non-spherical
models (see the left-hand panel of Fig. 7). This evolution differs
from the colder and fractal non-spherical models in that the central
number density rises slowly and does not decrease by 1~Myr. Regardless
of this difference, we show in the following sections that
the specific evolution makes little difference for $f_{\text{VMS}}$ or $\bar{M}_{\text{VMS}}$, and
the parameters that affect the results the most are the initial central
density and the mass of the system. While the change in dynamics is
interesting in its own right, it appears not to be of particular importance
for the results of our work here. For that reason we explore a
wider range of parameters for spherical star clusters and emphasize
that this is likely a safe approximation when studying parameters
such as $f_{\text{VMS}}$ or $\bar{M}_{\text{VMS}}$.

\paragraph{Setup of spherical star clusters}
Observations of local clusters
as well as simulations \citep{Lada2003,Girichidis2011,Bate2012} show that star formation occurs deeply embedded
in molecular clouds and that stars tend to form in the central high-density
regions. To set the radius of our spherical star clusters for a
given mass, we assume an isothermal density profile. We calculate
the radius of the cluster such that the mass enclosed equals $\epsilon M_{\text{clump}}$
and set this as the radius which encloses the stars up to a constant
factor $f_c$. The mass within this radius is redistributed into a Plummer
sphere and the remaining gas and dark matter are smoothed
over a second Plummer sphere with a radius of the original clump
multiplied by the same contraction factor $f_c$.

For an isothermal sphere,
\begin{equation}
\label{densstart}
M(r)=M_{\text{clump}}\frac{r}{R_{\text{clump}}},
\end{equation}
where $M_{\text{clump}}=M_*+M_g$ and $R_{\text{clump}}$ is the initial radius of the clump.  We can then calculate the stellar radius as,
\begin{equation}
R_*=\frac{M_*R_{\text{clump}}}{M_{\text{clump}}}=\epsilon R_{\text{clump}},
\end{equation}
which gives an enclosed mass profile for the stars,
\begin{equation}
M_*(r)=\frac{M_{*,p}r^3}{(r^2+a_{*,p}^2)^{\frac{3}{2}}}.
\end{equation}
Here, $M_{*,p}=[(\frac{3\pi}{16}f_c)^2+1]^{3/2}M_*$, $a_{*,p}=\frac{3\pi}{16}f_cR_*$ and $f_c$ is an arbitrary contraction factor.

We set the radius of the gas, $R_g$, equal to the original radius of the clump times $f_c$, and smooth the mass over a Plummer sphere, so we find that the enclosed mass profile for the gas is,
\begin{equation}
\label{densfin}
M_g(r)=\frac{M_{g,p}r^3}{(r^2+a_{g,p}^2)^{\frac{3}{2}}},
\end{equation}  
where $M_{g,p}=[(\frac{3\pi}{16}f_c)^2+1]^{3/2}M_g$ and $a_{g,p}=\frac{3\pi}{16}f_cR_g$.

Fig. 4 shows that a small amount of dark matter is also present
within the collapsing clump of gas at the centre of our mini-halo.
We include this mass in the external gas potential of our star cluster

In Fig. 8, we plot the average initial density profiles for spherical
clusters with $f_c=$~0.1, 0.2, 0.3, and 0.5 and compare with the
spherically averaged density profiles of the Plummer and constant
density models for our non-spherical star clusters discussed in Section
3.2.1. Note that the Plummer models used in the non-spherical
simulations are similar to the models used in the spherical simulations
with $f_c=0.5$. The constant-density models used in the
non-spherical simulations are less dense in the inner regions than
all of the spherical models considered here; however, they maintain
a higher density at larger radii. We found, however, that the
difference between the Plummer and constant-density models does
not significantly affect the probabilities of forming a VMS nor how
massive these stars become. For this reason, we only investigate
spherical clusters with Plummer density profiles.

\begin{figure}
\centerline{\epsfig{figure=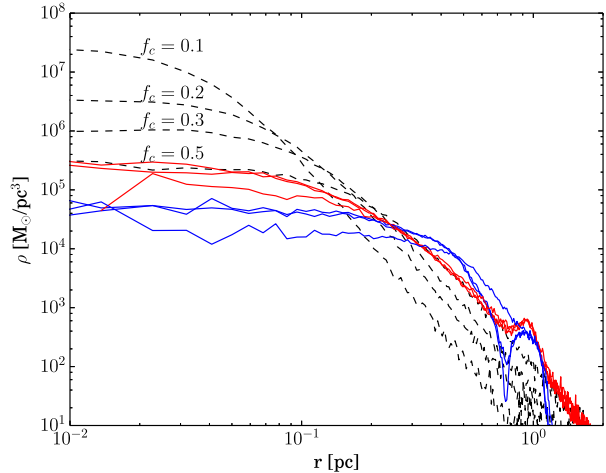,scale=.45}}
\caption{Density profiles of spherical models (dashed black) are compared with the spherically averaged density profiles of the non-spherical models.  The red line indicates models with a Plummer density profile and the blue line is for models with a constant density profile and the three different lines show $D=$3.0, 2.6, and 1.6.  The models with $D=$ 3.0 and 2.6 are indistinguishable while the models with $D=1.6$ have a slightly lower density in the inner regions of the cluster.  Spherical models with decreasing central densities represent clusters with $f_c=$ 0.1, 0.2, 0.3, and 0.5 as annotated.  $f_c$ is the fraction by which the initial radius of the clump is contracted to produce these models.}
\label{densinit}
\end{figure}

\begin{table*}
\centering
\begin{tabular}{@{}lccccccccccc@{}}
\multicolumn{12}{c}{{\bf Spherical Models}} \\
\hline
\multicolumn{4}{l}{{\bf Initial Conditions}} & \multicolumn{8}{l}{{\bf Results}} \\
\hline
$R_{*,vir}$&$f_c$&$Q$&$D$&$\bar{N}_{\text{coll}}$&$f_{\text{VMS}}$&$f_{\text{PISN}}$&$f_{\text{NE}}$&$\bar{M}_{\text{seed}}$&$M_{\text{VMS,max}}$&$\bar{M}_{\text{VMS}}$&$M_{\text{SMBH}}$\\
pc&&&&&per cent&per cent&per cent&M$_{\odot}$&M$_{\odot}$&M$_{\odot}$&$10^9$~M$_{\odot}$\\
\hline
0.071 & 0.1 & 0.5 & 3.0 & 9.08 & $82.0\pm12.8$ & $18.0\pm6.0$ & $0.0\pm0.0$ & 83.9 & 694.5 & 477.9 & 4.80\\  
0.143 & 0.2 & 0.5 & 3.0 & 4.56 & $58.0\pm10.8$ & $30.0\pm7.7$ & $12.0\pm4.9$ & 86.6 & 494.0  & 355.6 & 3.57\\ 
0.215 & 0.3 & 0.5 & 3.0 & 2.48 & $20.0\pm6.3$ & $55.0\pm10.5$ & $24.0\pm6.9$ & 86.3 & 408.8 & 336.5 & 3.38\\  
0.358 & 0.5 & 0.5 & 3.0 & 1.12 & $4.0\pm2.8$ & $38.0\pm8.7$ & $58.0\pm10.8$ & 85.3 & 264.1 & 263.6 & 2.65\\  
0.358 & 0.5 & 0.5 & 2.6 & 1.48 & $4.0\pm2.8$ & $44.0\pm9.4$ & $52.0\pm10.2$ & 92.4 & 382.4 & 338.2 & 3.39\\ 
0.358 & 0.5 & 0.5 & 1.6 & 3.28 & $4.0\pm 2.8$ & $50.0\pm 10.0$ & $46.0\pm9.6$ & 78.5 & 301.0 & 297.0 & 2.98\\ 
0.358 & 0.5 & 0.3 & 3.0 & 0.92 & $0.0\pm0.0$ & $40.0\pm8.9$ & $60.0\pm11.0$ & - & - & - & -\\
0.358 & 0.5 & 0.3 & 2.6 & 1.28 & $0.0\pm0.0$ & $52.0\pm10.2$ & $48.0\pm9.9$ & - & - & - & -\\  
0.358 & 0.5 & 0.3 & 1.6 & 3.02 & $6.0\pm3.5$ & $38.0\pm8.7$ & $56.0\pm10.6$ & 84.6 & 303.1 & 326.3 & 3.27\\ 
\hline
\end{tabular}
\caption{All models assume a TH\_Salp IMF with M$_{\text{min}}=1$~M$_{\odot}$, M$_{\text{max}}=100$~M$_{\odot}$, $\epsilon=2/3$, and $M_*=1.01\times10^4$~M$_{\odot}$.  $R_{*,vir}$: virial radius of the stellar component.  $f_c$: contraction factor.  For all models, 50 realizations are used.  See Table \ref{tab:nonideal} for definitions of other values.}
\label{tab:NB6ext}
\end{table*}

\paragraph{Varying $f_c$, $Q$, and $D$} 
In Table 2, we list the initial conditions
for these spherical models where we vary $f_c$, $Q$, and $D$. As we have
seen in Fig. 8, $f_c$ controls the initial central density of the cluster and
we found that for fixed mass, the initial central density is the most
important parameter in determining what percentage of clusters
produce a VMS as well as how massive the VMS can grow (see
Table 2). In Fig. 9, we plot the mass of the most massive VMS that
formed as well as the fraction of clusters which produced a VMS
as a function of initial central density. Both the fraction and the
mass of the VMS are increasing with the increase in initial central
density. The increase of both of these values is reasonably linear in
$\log \rho$ and can be approximated as
\begin{equation}
\label{eqn:fvms}
f_{\text{VMS}}=42.6\log_{10}(\rho_{*,max})-231.2 \text{ per cent},
\end{equation}
and
\begin{equation}
\label{eqn:mvms}
M_{\text{VMS,max}}=214.1\log_{10}(\rho_{*,max})-900.1 \text{ M}_{\odot}.
\end{equation}

\begin{figure}
\centerline{\epsfig{figure=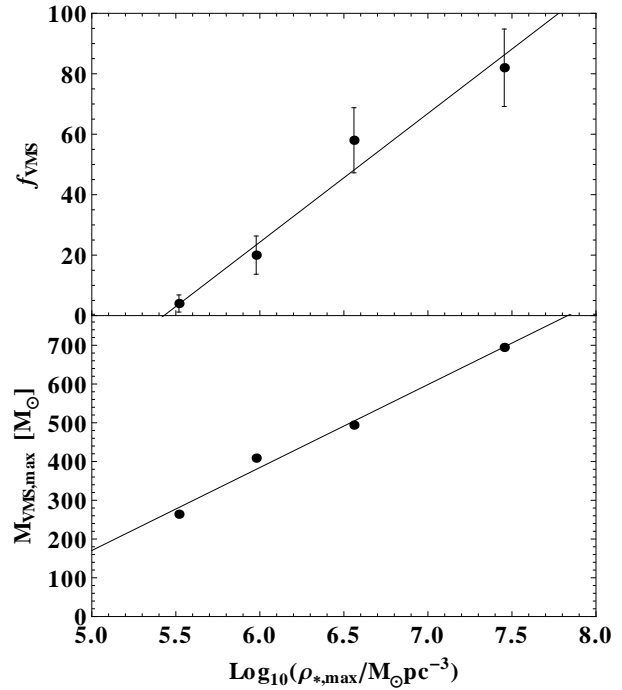,scale=.5}}
\caption{$f_{\text{VMS}}$ (top) and $M_{\text{VMS,max}}$ (bottom) as a function of the initial central stellar mass density for the TH\_Salp IMF assuming $S=0$ and $b=0$.  The data points come directly from the simulations and error bars are $1\sigma$ Poisson errors.  We over plot the best fit lines from equations (\ref{eqn:fvms}) \& (\ref{eqn:mvms}). }
\label{fmfit}
\end{figure}

Decreasing $Q$ from 0.5 to 0.3 does not affect $f_{\text{VMS}}$ or $\bar{N}_{\text{coll}}$, similarly
to the non-spherical models. We found a small increase in
$\bar{N}_{\text{coll}}$ for the more inhomogeneous models similar to what we found
for our non-spherical {\small NBODY6} simulations. This is likely to be due
to the higher local initial densities. The value of $D$, however, does
not change $f_{\text{VMS}}$. Based on the results in Table 2, a spherical cluster
with $f_c$ between 0.5 and 0.3, $Q=0.5$ and $D=3.0$ should give very
similar results to the non-spherical models.

\paragraph{Varying the IMF}
As the stellar IMF is highly uncertain,
especially at high redshift, we test the effect of varying the IMF.
The parameters of the different IMFs implemented are listed in
Table 3. In Table 4, we list the initial conditions and results of the
models where the IMF is varied. For these models, we have assumed
$f_c=0.2$ which was previously determined to have a high percentage
of models which produce a VMS (see Table 2).

\begin{table}
\centering
\begin{tabular}{@{}lccc@{}}
\multicolumn{4}{c}{{\bf Initial Mass Functions}} \\
\hline
Name & $M_{\text{min}}$ & $M_{\text{max}}$ & $\alpha$ \\
 & M$_{\odot}$ & M$_{\odot}$ & \\
\hline
Salp & 0.1 & 100.0 & $-2.35$ \\
\hline
TH\_Salp & 1.0 & 100.0 & $-2.35$ \\
\hline
\multirow{2}{*}{Kroupa} & 0.1 & 0.5 & $-1.3$ \\ 
& 0.5 & 100.0 & $-2.30$ \\
\hline
Flat & 1.0 & 100.0 & $-2.00$ \\
\hline
\end{tabular}
\caption{Table of the different IMFs sampled for the simulations of spherical star clusters.}
\label{tab:IMFs}
\end{table}

As the mean mass of the IMF is increased, the fraction of realizations
which produce a VMS remain consistent within error bars and
therefore, regardless of the IMF, the likelihood of producing a VMS
in a high-redshift NSC is the same for an NSC of this mass. The
average number of stellar collisions increases, however, for models
with lower average stellar mass ($\bar{m}$) due to the higher number density
of stars. Furthermore, $M_{\text{VMS,max}}$ and $\bar{M}_{\text{seed}}$ tend to increase as
the IMF becomes more top heavy. This is expected because there
are simply more high-mass stars in the runs with more top-heavy
IMFs and therefore the mean mass per collision will increase. We
can also see that $\bar{M}_{\text{seed}}$ slightly increases with increasing $\bar{m}$ which
also reflects the availability of more high-mass stars.

\paragraph{Primordial binaries and mass segregation:}
Observations
suggest the presence of a significant fraction of binaries in young
stellar clusters, especially for high-mass stars \citep{Hut1992,Sana2009}. A large binary fraction at the centres of star
clusters can effectively heat the cluster and prevent core collapse.
This decreases the maximum value of the central density which
may inhibit the formation of a VMS due to collisional runaway.
When clusters are initialized with binaries, massive stars with mass
greater than $5$~M$_{\odot}$ are preferentially put in binaries. For low-mass
binaries, we use a period distribution consistent with \cite{Kroupa1995}
and for binary stars which have a primary star with mass greater
than $5$~M$_{\odot}$, we adopt the period distribution consistent with \cite{Sana2011}. We test the effects of including primordial binaries
as well as primordial mass segregation for these spherical clusters
and list the initial conditions in Table 5.

\begin{table*}
\centering
\begin{tabular}{@{}lcccccccccc@{}}
\multicolumn{11}{c}{{\bf Spherical Models: Varying the Stellar IMF}} \\
\hline
\multicolumn{3}{l}{{\bf Initial Conditions}} & \multicolumn{8}{l}{{\bf Results}} \\
\hline
IMF & $\bar{M}_{\text{max}}$ & $\bar{m}$ & $\bar{N}_{\text{coll}}$ & $f_{\text{VMS}}$ & $f_{\text{PISN}}$ & $f_{\text{NE}}$ & $\bar{M}_{\text{seed}}$ & $M_{\text{VMS,max}}$ & $\bar{M}_{\text{VMS}}$ & $M_{\text{SMBH}}$ \\
& M$_{\odot}$ & M$_{\odot}$ & & per cent & per cent & per cent & M$_{\odot}$ & M$_{\odot}$ & M$_{\odot}$ & $10^9$~M$_{\odot}$ \\
\hline
Salp & 79.5 & 0.35 & 8.95 & $40.0\pm14.1$ & $50.0\pm15.8$ & $10.0\pm7.1$ & 75.2 & 353.1 & 295.3 & 2.96 \\
TH\_Salp & 90.4 & 3.09 & 4.56 & $58.0\pm10.8$ & $30.0\pm7.7$ & $12.0\pm4.9$ & 86.6 & 494.0  & 355.6 & 3.57\\
Kroupa & 88.8 & 0.64 & 5.80 & $44.0\pm9.4$ & $38.0\pm8.7$ & $18.0\pm6.0$ & 84.0 & 433.2 & 333.9 & 3.35\\
Flat & 96.1 & 4.65 & 3.44 & $48.0\pm9.8$ & $40.0\pm8.9$ & $12.0\pm4.9$ & 89.9 & 591.5 & 376.7 & 3.78\\
\hline
\end{tabular}
\caption{$\bar{M}_{\text{max}}$:  the average maximum mass star over all realizations of the cluster.  $\bar{m}$: the average stellar mass over all realizations of the cluster.  See Table \ref{tab:nonideal} for definitions of other values.  50 realizations were run for all models except the model with the Salp IMF where only 20 simulations are used.}
\label{tab:NB6res}
\end{table*}

\begin{table*}
\centering
\begin{tabular}{@{}lcccccccccccc@{}}
\multicolumn{13}{c}{{\bf Spherical Models: Primordial Binaries and Mass Segregation}} \\
\hline
\multicolumn{5}{l}{{\bf Initial Conditions}} & \multicolumn{8}{l}{{\bf Results}} \\
\hline
IMF & $\bar{M}_{\text{max}}$ & $\bar{m}$ & $S$ & $b$ & $\bar{N}_{\text{coll}}$ & $f_{\text{VMS}}$ & $f_{\text{PISN}}$ & $f_{\text{NE}}$ & $\bar{M}_{\text{seed}}$ & $M_{\text{VMS,max}}$ & $\bar{M}_{\text{VMS}}$ & $M_{\text{SMBH}}$ \\
& M$_{\odot}$ & M$_{\odot}$ & & & & per cent & per cent & per cent & M$_{\odot}$ & M$_{\odot}$ & M$_{\odot}$ & $10^9$~M$_{\odot}$ \\
\hline
TH\_Salp & 90.4 & 3.09 & 0.0 & 0.0 & 4.56 & $58.0\pm10.8$ & $30.0\pm7.7$ & $12.0\pm4.9$ & 86.6 & 494.0  & 355.6 & 3.57\\
TH\_Salp & 90.4 & 3.11 & 0.5 & 0.0 & 4.78 & $66.0\pm11.4$ & $26.0\pm7.2$ & $8.0\pm4.0$ & 85.8 & 516.8 & 352.7 & 3.54\\
TH\_Salp & 90.6 & 3.08 & 1.0 & 0.0 & 3.64 & $36.0\pm8.5$ & $48.0\pm9.8$ & $16.0\pm5.7$ & 83.2 & 406.1 & 318.5 & 3.20\\
TH\_Salp & 92.6 & 3.10 & 0.0 & 0.5 & 17.6 & $32.0\pm8.0$ & $50.0\pm10.0$ & $18.0\pm6.0$ & 81.9 & 438.4 & 319.0 & 3.20\\
TH\_Salp & 91.5 & 3.09 & 0.0 & 1.0 & 22.0 & $22.0\pm6.6$ & $50.0\pm10.0$ & $28.0\pm7.5$ & 79.0 & 445.0 & 327.0 & 3.28\\
TH\_Salp & 91.4 & 3.09 & 0.5 & 0.5 & 18.0 & $42.0\pm9.2$ & $44.0\pm9.4$ & $14.0\pm5.3$ & 81.5 & 515.1 & 344.0 & 3.45\\
TH\_Salp & 90.6 & 3.09 & 1.0 & 0.5 & 16.8 & $24.0\pm6.9$ & $36.0\pm8.5$ & $40.0\pm8.9$ & 83.2 & 477.7 & 336.7 & 3.38\\
TH\_Salp & 90.0 & 3.09 & 0.5 & 1.0 & 23.7 & $46.0\pm9.6$ & $36.0\pm8.5$ & $18.0\pm6.0$ & 80.1 & 465.3 & 353.7 & 3.55\\
TH\_Salp & 91.2 & 3.08 & 1.0 & 1.0 & 22.7 & $26.0\pm7.2$ & $38.0\pm8.7$ & $26.0\pm7.2$ & 82.2 & 595.6 & 365.9 & 3.67\\
Kroupa & 88.8 & 0.64 & 0.0 & 0.0 & 5.80 & $44.0\pm9.4$ & $38.0\pm8.7$ & $18.0\pm6.0$ & 84.0 & 433.2 & 333.9 & 3.35\\
Kroupa & 87.2 & 0.64 & 0.5 & 0.0 & 6.18 & $42.0\pm9.2$ & $50.0\pm10.0$ & $8.0\pm4.0$ & 80.0 & 410.7 & 336.6 & 3.38\\
Kroupa & 85.8 & 0.64 & 1.0 & 0.0 & 5.24 & $52.0\pm10.2$ & $38.0\pm8.7$ & $10.0\pm4.5$ & 78.9 & 435.5 & 327.7 & 3.29\\
\hline
\end{tabular}
\caption{Note that the TH\_Salp and Kroupa models with $S=0$ and $b=0$ are the same models listed in Table \ref{tab:NB6res} and are shown here for ease of comparison.  See Table \ref{tab:nonideal} for definitions of values.  For all models, 50 realizations are run.}
\label{tab:NB6resSB}
\end{table*}

Introducing mass segregation for models with TH\_Salp and
Kroupa IMFs makes no difference to $\bar{M}_{\text{VMS}}$. The same is true when
binaries are introduced. In all cases, the average VMS undergoes
$\approx1-3$ collisions after the PISN mass threshold and with these low
number statistics, it is unsurprising that $\bar{M}_{\text{VMS}}$ is roughly the same
regardless of variations to these initial parameters. We do, however,
see a drastic increase in $\bar{N}_{\text{coll}}$ when the clusters are initialized with
large binary fractions. The increase of the total number of collisions
in the cluster is clearly not affecting $\bar{M}_{\text{VMS}}$. For stars that
are in binaries from the beginning, interactions with other stars can
efficiently remove energy from the binary and eventually drive the
binary to a merger. Without binaries, the massive stars first have
to sink to the centre of the cluster and dynamically form binaries
before merging can take place. The average mass of the mergers is,
however, much lower for simulations with primordial binaries and
these newly formed, merged stars tend to have very little impact on
the formation of a VMS. Despite the evolution of the star cluster
being significantly different for simulations with primordial mass
segregation and binaries, we can make robust predictions for the
average mass of a VMS that will form.

Although $\bar{M}_{\text{VMS}}$ is only weakly dependent on most assumptions
we have made, we do find a factor of 3 change in the fraction of
models which produce a VMS when including mass segregation
and binaries. While this does not affect the mass function of these
objects, it does affect their number density. Introducing binaries
tends to decrease the number of VMSs which form while no trend
is evident for increasing the primordial mass segregation. Including
the binaries tends to heat the core and eject the low-mass stars,
and this influences the evolution of the core of the cluster where
the collisions occur. The effect of changing these assumptions is,
however, much less significant than changes to the central mass
density of the star cluster.

\paragraph{Varying the mass of the NSC}
Thus far, our work has focused
on one simulation of one mini-halo with one NSC with a
particular mass motivated by our cosmological zoom-in simulation.
The Universe likely exhibits a range of NSC masses with varying
initial central densities and we also want to explore how our results
depend on the assumed mass of the NSC. Although the first collapsing
halo in our simulation is not be metal enriched, to get an
idea of the variance in the mass of the NSCs in the early Universe,
we can apply the same criteria that we used to define the NSC in
the secondary collapsing halo to determine an NSC mass for the
first halo. The total mass of the NSC in this halo including gas and
dark matter is $\approx2.3\times10^4$~M$_{\odot}$, which is clearly more massive than
the NSC in the secondary collapsing object. We run a few additional
simulations where we multiply the total mass of the spherical
$f_c=0.2$ model by some factor $f_m$ and correspondingly increase the
radius of this model by $f_m^{1/3}$ in order to determine how massive an
IMBH may become for different NSC masses. In Table 6, we list
the initial conditions and results of these models which have been
averaged over 25 realizations. For all models, we assume a TH\_Salp
IMF and Plummer model initial conditions\footnote{Note that models with larger $f_m$ maintain a higher average density at larger radii.}.

\begin{table*}
\centering
\begin{tabular}{@{}lcccccccccccc@{}}
\multicolumn{13}{c}{{\bf Spherical Models: Varying the Mass of the NSC}} \\
\hline
\multicolumn{5}{l}{{\bf Initial Conditions}} & \multicolumn{8}{l}{{\bf Results}} \\
\hline
$f_m$ & $\bar{M}_{\text{max}}$ & $\bar{m}$ & $S$ & $b$ & $\bar{N}_{\text{coll}}$ & $f_{\text{VMS}}$ & $f_{\text{PISN}}$ & $f_{\text{NE}}$ & $\bar{M}_{\text{seed}}$ & $M_{\text{VMS,max}}$ & $\bar{M}_{\text{VMS}}$ & $M_{\text{SMBH}}$ \\
& M$_{\odot}$ & M$_{\odot}$ & & & & per cent & per cent & per cent & M$_{\odot}$ & M$_{\odot}$ & M$_{\odot}$ & $10^9$~M$_{\odot}$ \\
\hline
2 & 94.8 & 3.09 & 0.0 & 0.0 & 7.20 & $80.0\pm 17.9$ & $20.0\pm 8.9$ & $0.00\pm 0.0$ & 89.1 & 875.8 & 495.2 & 4.97 \\
3 & 96.7 & 3.09 & 0.0 & 0.0 & 9.48 & $96.0\pm 19.6$ & $4.0\pm 4.0$ & $0.00\pm 0.0$ & 87.3 & 860.9 & 611.1 & 6.13 \\
4 & 97.2 & 3.09 & 0.0 & 0.0 & 11.36 & $96.0\pm 19.6$ & $4.0\pm 4.0$ & $0.00\pm 0.0$ & 88.2 & 1016.7 & 671.4 & 6.74 \\
\hline
\end{tabular}
\caption{$f_m$: fraction by which the mass of the NSC is scaled.  The radius is scaled by $f_m^{1/3}$.  See Table \ref{tab:nonideal} for definitions of other values.}
\label{tab:changemass}
\end{table*}

It is clear that for these more massive NSCs, the majority are
likely to form an IMBH with some producing IMBHs with masses
greater than $1000$~M$_{\odot}$. The average masses of the VMS are also
significantly higher.

\subsection{Caveats and Limitations}
The initial conditions of our models, in particular the central number
and mass densities, play a major role in determining the fraction
of clusters which produce a VMS. Owning to numerical limitations,
to the best of our knowledge, there is no study to date which has
produced an embedded cluster such as the one presented in this
paper that had an IMF consistent with observations which has been
evolved from the protostellar accretion phase all the way to the onset
of the ZAMS. There is thus a clear disconnect between the types of
simulations which can form protostellar cores directly from molecular
clouds and the direct N-body simulations which can follow the
subsequent evolution of these stars once the cluster has completely
formed. Simulations which follow the initial formation and gas accretion
on to protostellar cores explore how the physical properties
of the collapsing molecular cloud and the accretion dynamics affect
the initial conditions of the star cluster \citep{Bate1998,Bate2005,Bate2012,Krumholz2012}. We are unable
to capture this direct link between the properties of the birth cloud
and the initial conditions of the star cluster with our direct $N$-body
calculations. In order to properly model this system, we would have
to self-consistently predict the distribution function and IMF of stars
from the cosmological simulation and run the simulation far beyond
the initial stages of collapse. This is, unfortunately, beyond current
numerical capabilities. We have, however, mitigated this by studying
a large number of possible initial conditions for the star clusters
we have simulated. We believe that we can therefore confidently
conclude that for our choice of physically motivated and reasonable
assumptions, our simulations predict collisional runaway to be a
promising mechanism for the formation of VMSs.

We have allowed the stars to grow far beyond the masses at which
stellar evolution is well understood due to the lack of observational
constraints. In this regime, it is not obvious whether such an object
is stable and evolves like a normal star especially because it is
created by merging. For low-metallicity stars, the main-sequence
lifetime and radius, which are the most important stellar parameters
for this work, begin to flatten at high masses. For these reasons, we
have chosen to evolve them as 100~M$_{\odot}$ stars. Two other processes
along these lines are neglected in our simulations that will affect
stellar collisions, the inflation of the collisional remnant's radius
and the possible increase in the remnant's main-sequence lifetime.
When stars undergo collisions, some of the kinetic energy from
the collision is absorbed into the envelope of the primary which
can cause the merger remnant to have a much greater radius than
a comparable star with that mass which evolves normally on the
main sequence \citep{Dale2006}. The lifetime of this stage
is likely much shorter than the main-sequence lifetime of the star.
However, the gravitationally focusing cross-section of the star scales
with radius so the probability of undergoing a collision can greatly
increase during this period \citep{Dale2006}. Further collisions
will cause a similar effect, and thus, if the time between collisions
is short, the inflated radius can be sustained for long periods of
time. The mixing of the two colliding stars can introduce a fresh
source of hydrogen into the core and can increase the main-sequence
lifetime of the remnant compared to a star with a similar mass that
evolves normally on the main sequence \citep{Glebbeek2008,Glebbeek2013}. The prolonged lifetime will increase
the number of stellar collisions that remnant might undergo. Both
effects tend to improve the prospect of forming a VMS and by
neglecting them, our results should be conservative in this regard.

\section{Implications for the (early) growth of supermassive black holes}
\label{discuss}
We have demonstrated that high-redshift, dense, metal-poor NSCs
are likely to host runaway stellar collisions which produce VMSs
and that this process is robust to a wide variety of assumptions.
We have, however, not addressed here what happens to the VMS
once it forms and there are also no observational constraints for
stars of this mass. \cite{Heger2003} predict that VMSs with the
mass and metallicity as in our work end their lives by directly
collapsing into an IMBH with minimal mass-loss. However, these
predictions need probably to be considered with a healthy scepticism
given the lack of observational constraints. Furthermore, we
have assumed that when the stars merge, the remnant becomes an
ordinary main-sequence star. This is also uncertain as the impact
of the merge certainly disrupts the star at least temporarily and
the subsequent stellar evolution also remains uncertain. With these
caveats in mind, we now assess briefly whether this mechanism can
be responsible for producing the population of bright quasars at
$z\approx6-7$ which has an observed lower limit of $1.1\times10^{-9}$ Mpc$^{-3}$ \citep{Venemans2013}. To do this, we estimate the mass to which
IMBHs can grow by these lower redshifts and their expected space
density.

Under the assumption of Eddington-limited accretion, the mass
of a black hole increases as $M=M_0e^{((1-\epsilon)/\epsilon)(t/t_{\text{Edd}})}$ where $\epsilon$ is the radiative efficiency and $t_{\text{Edd}}=\sigma_Tc/(4\pi Gm_{\text p})$ \citep{Frank2002}. The size of $\epsilon$ is somewhat uncertain and depends on the
properties of the accretion disc which surrounds the black hole as
well as the spin of the black hole. Our direct N-body simulations
are run for 3.5~Myr after the point at which we extract the clump
from the simulation. This corresponds to the lifetimes of the most
massive stars. By this point, the hydrodynamic simulation would
have evolved to $z=25.88$. Assuming a canonical value of $\epsilon=0.1$ we calculate the expected mass of the black hole at z = 6 and tabulate
the results in Tables~1, 2, 4, 5, and 6. It is clear that regardless of
many of the initial assumptions of the IMF, binary fraction, initial
mass segregation, SFE, and density profile, the average masses of
the VMSs that do form are sufficient to grow to the masses of the
black holes powering high-redshift quasars that we observe at $z\gtrsim6$
if the black holes can be continuously fed at the Eddington accretion
rate.

The mass range that we predict for our VMSs is well within the
range predicted by some simulations of Pop. III stars \citep{Hirano2014}. For simple models assuming Eddington-limited accretion,
there is thus little difference between assuming that the growth of
SMBHs starts with the remnants of Pop. III stars and assuming that
the growth is seeded by the VMS resulting from runaway growth
in Pop. II star cluster. What makes the latter route perhaps more
promising is, however, the very different environment in which
the collisional runaway VMSs form compared to that of Pop. III
stars. \cite{Johnson2007} predict a long time delay for efficient accretion on to a black hole in the mass range studied in this
work due to radiative feedback on the surrounding medium. Only
in haloes where the gas density remains sufficiently high can this be
avoided although even if the gas remains at high densities, the radiative
feedback still may limit the accretion to sub-Eddington levels
\citep{Milosavljevic2009}. Mergers with surrounding
mini-haloes may provide the dense gas supply necessary to overcome
most of the radiative feedback and efficiently feed the black
hole. Recall that the formation of a Pop. II NSC at these very high
redshifts we study requires a close neighbouring halo which can
pollute the halo in which the IMBH forms with metals. The haloes
in our hydrodynamic simulations merge shortly after the IMBH is
likely to form. Thus, it appears unlikely that an IMBH which may
form in our simulations would suffer a long period of inefficient
accretion.

A further difference between a Pop. III remnant black hole and
the IMBHs forming from runaway stellar collisions is the presence
of a surrounding star cluster in the latter scenario. \cite{Alexander2014} discuss the evolution of low-mass black holes
with $\approx10$~M$_{\odot}$ embedded in a star cluster fed by dense gas flows and
accreting at super-Eddington rates due to random motions within
the cluster. If high-redshift, dense star clusters have very top-heavy
IMFs, this may indeed be relevant for more massive black hole
seeds. A similar mechanism was indeed suggested by \cite{Davies2011} where dense inflows of gas can initiate core
collapse and cause efficient merging in the central regions of the
cluster. The presence of the cluster after the seed has formed may
be key to allowing it to grow efficiently.

Assuming that these SMBH seeds can accrete at the required
rates, we can calculate an approximate upper limit on the number
density of SMBHs by calculating the total number density of
haloes enriched above the critical metallicity at the final redshift,
$z_{fin}$, such that IMBHs can grow to $10^9$~M$_{\odot}$ by $z=6$. Assuming
that the average mass of an IMBH that forms in our simulation is
300 M$_{\odot}$, $z_{fin}=20$.  The number density of SMBHs at $z=6$ then is
approximately
\begin{equation}
\label{eqn:grow}
n_{\text{SMBH}}=f_pf_ff_{\text{edd}}n_{gal}(>M_{\text{thresh}}),
\end{equation}
where $f_p$ is the fraction of haloes at $z_{fin}$ that have been metal enriched
above $Z_{\text{crit}}$, $f_f$ is the fraction of these haloes that form an
IMBH, $f_{\text{edd}}$ is the fraction of the black holes which can sustain
Eddington-limited accretion, and $n_{gal}(>M_{\text{thresh}})$ is the total number
density of galaxies above the mass threshold, $M_{\text{thresh}}$, which are
possible sites for forming an IMBH. We set $M_{\text{thresh}}=5\times10^6$~M$_{\odot}$,
and using the Jenkins mass function \citep{Jenkins2001}, we find, $n_{gal}\approx8$~Mpc$^{-3}$. For the mass of the NSC which formed in our
simulation, $f_f$ is likely  $<0.1$. There is little constraint on the value
of $f_p$ as we have little knowledge of the metal enrichment of the
intergalactic medium at these extremely high redshifts. Metal enrichment
in the early Universe is almost certainly very patchy and
confined to overdense regions. However, even for $f_f=0.01$, a rather small value of $f_p\approx10^{-5}$ would be sufficient to explain the observed
SMBH number density based on this simple estimate with $f_{\text{edd}}=1$.  The value of $f_{\text{edd}}$ also remains highly uncertain and it is unlikely
to be unity given the plethora of environmental feedback mechanisms
that inhibit efficient accretion on to black holes \citep{Johnson2007,Milosavljevic2009,Park2011}. For
this to occur, black holes would require a constant supply of cold,
low-angular-momentum gas, and since we do not follow the hydrodynamical
simulation past the formation of the NSC birth cloud, it
is uncertain whether such a reservoir is available. Note, however,
that the mass function of IMBHs forming in high-redshift NSCs
is unlikely to be a delta function at the minimum mass required to
form an SMBH by $z=6$ so $f_{\text{edd}}$ need not be 1 for this process to produce
the population of observed SMBHs at $z=6$. In addition, other
mechanisms may allow black holes to accrete at super-Eddington
rates when a surrounding star cluster is present \citep{Davies2011,Alexander2014}.

\section{Conclusions}
Using a combination of high-resolution, hydrodynamic cosmological
zoom-in simulations and spherical and non-spherical direct
N-body simulations, we have demonstrated that stellar runaway
growth at the centre of nuclear Pop. II star clusters in high-redshift
protogalaxies ($20\lesssim z\lesssim30$) metal enriched by nearby companions
is a promising route to form VMSs with masses as high as $300-1000$~M$_{\odot}$. We find that the average masses of the VMSs that are
produced in NSCs with an expected total mass of typically $M_*\approx10^4$~M$_{\odot}$ are relatively robust to changes in the stellar IMF, number
of primordial binaries, initial degree of mass segregation, as well
as initial density profile, but increase strongly with the increase in
initial central density and total mass of the star cluster. If the VMSs
formed in this way can directly collapse to IMBHs with moderate
mass-loss, they are promising seeds for growth into the billion
solar-mass black holes observed at $z\approx6$. Our simulations further
predict an enhanced number of PISNs as the simulations which fail
to produce a VMS often host at least one or two high-mass collisions.
This may result in a rapid early enrichment of the IGM with
metals and may cause the early pollution of other protogalaxies
making them also susceptible to the collisional runaway process.
Modelling the evolution of the host galaxy post-supernova will be
needed to predict the effects of these PISNs on their environment

The presence of a large numbers of accreting IMBHs as well
as an enhanced rate of PISN should significantly alter the early
evolution of galaxies and will have important implications for the
interpretation of observations of the high-redshift Universe.

\section*{Acknowledgements}
We thank the referee for their comments and revisions. HK is grateful
to Sverre Aarseth and Simon Karl for useful discussions about
{\small NBODY6} and very much appreciates the hospitality of the Foundation
Boustany during his stay in Monaco where parts of this manuscript
were written. We also thank John Regan, Chris Tout, and Sverre
Aarseth for comments which greatly improved the manuscript. This
work made considerable use of the open source analysis software
{\small PYNBODY} \citep{pynbody}. HK's work is partially supported
by Foundation Boustany, Cambridge Overseas Trust, and an Isaac
Newton Studentship. Support by ERC Advanced Grant 320596
'The Emergence of Structure during the Epoch of Reionization' is
gratefully acknowledged.

This work was performed using the DiRAC/Darwin Supercomputer
hosted by the University of Cambridge High Performance
Computing Service (http://www.hpc.cam.ac.uk/), provided by Dell
Inc. using Strategic Research Infrastructure Funding from the
Higher Education Funding Council for England and funding from
the Science and Technology Facilities Council. Cosmological hydrodynamic
simulations were performed on the DARWIN cluster
while all direct $N$-body runs were performed on the WILKES

Furthermore, this work used the DiRAC Complexity system,
operated by the University of Leicester IT Services, which
forms part of the STFC DiRAC HPC Facility (www.dirac.ac.uk).
This equipment is funded by BIS National E-Infrastructure capital
grant ST/K000373/1 and STFC DiRAC Operations grant
ST/K0003259/1. DiRAC is part of the National E-Infrastructure.

\bibliographystyle{./mn2e}
\bibliography{./HkDsMH_2015}

\appendix
\section{Convergence Tests for Non-Spherical Initial Conditions}
\label{converge}
In Fig. 3, we have shown that despite the fragmentation present
in higher resolution simulations, the mass contained in the central
region of the central collapsing galaxy remains well converged. To
improve on this point, we plot the phase space diagrams of density
versus temperature at three different times for the $l_{\text{max}}=20$ simulation
in Fig. A1. The evolution is nearly identical to the $l_{\text{max}}=19$
run which is shown in Fig. 2, and the main difference between these
two simulations is caused by clump-clump interactions which affect
the central structure. The structure of the cells we identify as star
forming in the $l_{\text{max}}=20$ simulation changes and becomes denser
compared to the $l_{\text{max}}=19$ run (see Fig. 5). We can test how these
denser clumps affect the formation of VMSs by applying the same
method for creating initial conditions as used in Section 3.1.1 to the
highest resolution simulation.

\begin{figure*}
\centerline{\includegraphics[scale=1.1,angle=-90,clip,trim=7.5cm 1.8cm 7.5cm 9.4cm]{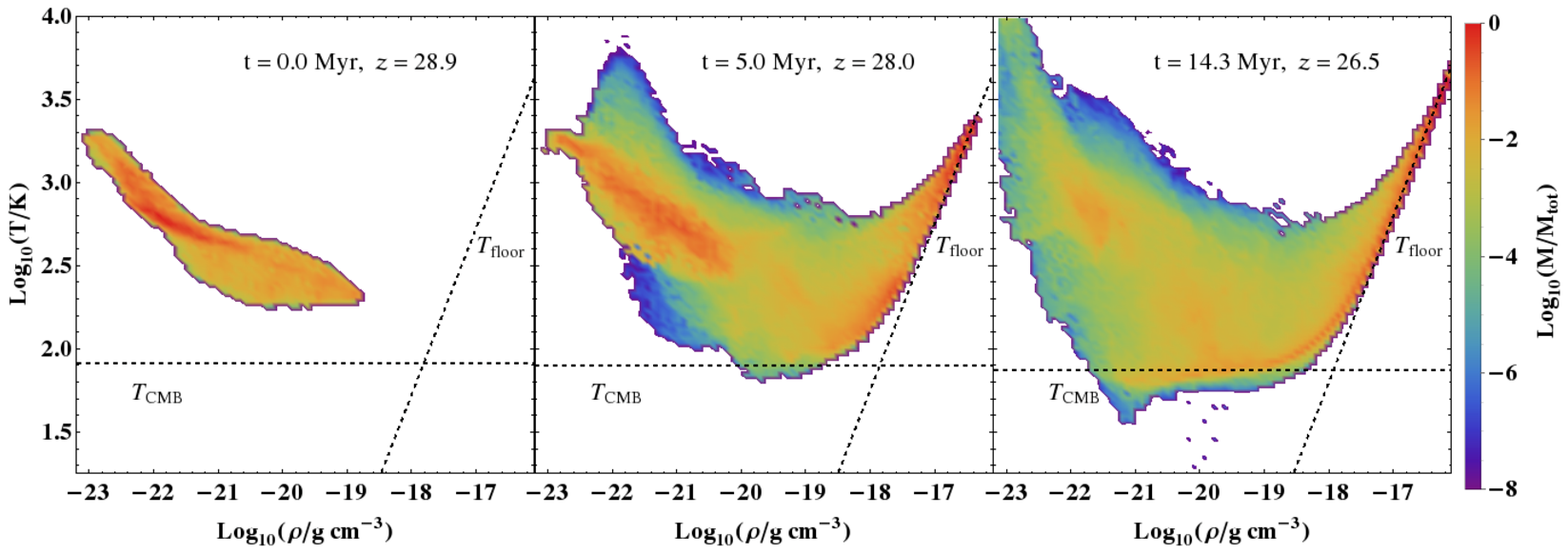}}
\caption{Mass-weighted phase space diagrams of density versus temperature at the initial collapse ({\it left}), 5 Myr after the collapse ({\it middle}), and 14.3 Myr after collapse ({\it right}).  This includes all gas within 10 pc of the densest cell.  The gas cools to a few hundred kelvin initially due to H$_2$ cooling then reaches the CMB temperature floor due to HD and metal cooling.  The dashed lines represent the CMB temperature floor and the artificial temperature floor as labelled.}
\label{PT:BA20}
\end{figure*}

In Table A1, we list the initial conditions and results of direct $N$-body
simulation run using as input the $l_{\text{max}}=20$ run. Even though
these runs have slightly less mass, we find that $f_{\text{VMS}}$ has increased
due to the increase in density. Overall, $\bar{M}_{\text{VMS}}$ remains reasonably
consistent with our previous results which suggest that this mass is
robust.

\begin{table*}
\centering
\begin{tabular}{@{}lcccccccccccc@{}}
\multicolumn{12}{c}{{\bf Non-Spherical Models: Convergence}} \\
\hline
\multicolumn{5}{l}{{\bf Initial Conditions}} & \multicolumn{8}{l}{{\bf Results}} \\
\hline
$\rho_{\text{init}}$ & IMF & $Q$ & $D$ & $S$ &$\bar{N}_{\text{coll}}$&$f_{\text{VMS}}$&$f_{\text{PISN}}$&$f_{\text{NE}}$&$\bar{M}_{\text{seed}}$&$M_{\text{VMS,max}}$&$\bar{M}_{\text{VMS}}$&$M_{\text{SMBH}}$\\
& & & & & & & & & M$_{\odot}$ & M$_{\odot}$ & M$_{\odot}$ & 10$^9$~M$_{\odot}$ \\
\hline
CD & TH\_Salp & 0.5 & 3.0 & 0.0 & 2.38 & $16.0\pm5.7$ & $46.0\pm9.6$ & $38.0\pm8.7$ & 87.6 & 311.8 & 290.4 & 2.91\\
CD & TH\_Salp & 0.3 & 3.0 & 0.0 & 2.40 & $18.0\pm6.0$ & $56.0\pm10.6$ & $26.0\pm7.2$ & 84.8 & 351.3 & 294.3 & 2.95\\
CD & TH\_Salp & 0.1 & 3.0 & 0.0 & 2.44 & $24.0\pm6.9$ & $32.0\pm8.0$ & $44.0\pm9.4$ & 89.0 & 349.0 & 296.6 & 2.98\\
\hline
\end{tabular}
\caption{50 Realizations of each model are generated.  See Table \ref{tab:nonideal} for definitions of the listed quantities.}
\label{tab:nonidealL20}
\end{table*}

\section{Varying the Star Formation Efficiency}
\label{vsfe}
\begin{table*}
\centering
\begin{tabular}{@{}lccccccccccccc@{}}
\multicolumn{14}{c}{{\bf Spherical Models: Varying the Star Formation Efficiency}} \\
\hline
\multicolumn{6}{l}{{\bf Initial Conditions}} & \multicolumn{8}{l}{{\bf Results}} \\
\hline
$\epsilon$&$M_*$&$R_{*,vir}$&$f_c$&$Q$&$D$&$\bar{N}_{\text{coll}}$&$f_{\text{VMS}}$&$f_{\text{PISN}}$&$f_{\text{NE}}$&$\bar{M}_{\text{seed}}$&$M_{\text{VMS,max}}$&$\bar{M}_{\text{VMS}}$&$M_{\text{SMBH}}$\\
&$10^4$~M$_{\odot}$&pc&&&&&per cent&per cent&per cent&M$_{\odot}$&M$_{\odot}$&M$_{\odot}$&$10^9$~M$_{\odot}$\\
\hline
1/2 & 0.76 & 0.107 & 0.2 & 0.5 & 3.0 & 3.90 & $50.0\pm10.0$ & $40.0\pm8.9$ & $10.0\pm4.5$ & 84.1 & 492.8 & 345.8 & 3.47\\ 
1 & 1.52 & 0.215 & 0.2 & 0.5 & 3.0 & 4.40 & $45.0\pm9.5$ & $40.0\pm8.9$ & $15.0\pm5.5$ & 82.8 & 551.0 & 402.5 & 4.04\\
\hline
\end{tabular}
\caption{20 Realizations of each model are generated.  See Table \ref{tab:nonideal} for definitions of the listed quantities.}
\label{tab:NB6sfe}
\end{table*}

Most of our models were based on the assumption of an SFE of $\epsilon=2/3$  which is motivated by a series of observations and simulations
of relevant environments. We test two models where the
SFE is changed and these models are listed in Table B1. The cluster
parameters are derived using equations (4)-(7) and note that the
initial central densities of the clusters are dependent on the chosen $\epsilon$. We see that VMSs can form in clusters with lower $M_*$ although
they likely require higher central densities.

\section{Alternative Model for Mass Segregation}
\label{subr}

\begin{table*}
\centering
\begin{tabular}{@{}lccccccccc@{}}
\multicolumn{10}{c}{{\bf Spherical Models: Alternative Mass Segregation}} \\
\hline
\multicolumn{2}{l}{{\bf Initial Conditions}} & \multicolumn{8}{l}{{\bf Results}} \\
\hline
Profile&Parameters&$\bar{N}_{\text{coll}}$&$f_{\text{VMS}}$&$f_{\text{PISN}}$&$f_{\text{NE}}$&$\bar{M}_{\text{seed}}$&$M_{\text{VMS,max}}$&$\bar{M}_{\text{VMS}}$&$M_{\text{SMBH}}$\\
&&&per cent&per cent&per cent&M$_{\odot}$&M$_{\odot}$&M$_{\odot}$&$10^9$~M$_{\odot}$\\
\hline
Subr & $S=0.5$ & 3.92 & $50.0\pm10.0$ & $36.\pm8.5$0 & $14.0\pm5.3$ & 84.4 & 559.5 & 350.3 & 3.51\\
\hline
\end{tabular}
\caption{See Table \ref{tab:nonideal} for definitions of the listed quantities.}
\label{tab:NB6dp}
\end{table*}

We generate one additional model in order to compare an alternative
method for primordial mass segregation with the one we have used
in the text in order to determine how sensitive our results are to the
algorithm used. Here we adopt the algorithm of \cite{Subr2008} and the results are listed in Table C1. For this
model, we use the TH\_Salp IMF and it is directly comparable with
the model in Section 3.2 with the same IMF and initial degree of primordial
mass segregation. Within the 1$\sigma$ errors on the percentages
of VMSs and PISNs that form, these models are entirely consistent
suggesting that our choice of algorithm is not significantly affecting
the results.

\label{lastpage}
\end{document}